\documentclass[twocolumn,showpacs,superscriptaddress,prb,aps,floatfix]{revtex4-2}
\usepackage{graphicx}
\usepackage{rotating}
\usepackage{amsmath}
\usepackage[usenames,dvipsnames]{color}
\usepackage{float}
\usepackage{rotating}
\usepackage{booktabs}
\usepackage{multirow}
\usepackage{longtable}
\usepackage[colorlinks,linkcolor=blue,citecolor=blue]{hyperref}
\usepackage[T1]{fontenc}
\usepackage{bm}
\usepackage{upgreek}
\usepackage{comment}
\makeatletter

\newcommand{\Rmnum}[1]{\expandafter\@slowromancap\romannumeral #1@}

\newcommand{\DC}[1]{\textcolor{red}{To Domenico and Cesare:}}

\makeatletter

\hfuzz=\maxdimen
\begin{document}

\title{Hierarchy of Exchange-Correlation Functionals in Computing Lattice Thermal Conductivities of Rocksalt and Zincblende Semiconductors}

\author{Jiacheng Wei}
\affiliation{Key Laboratory of Advanced Materials and Devices for Post-Moore Chips, Ministry of Education, University of Science and Technology Beijing, Beijing 100083, China}
\affiliation{School of Mathematics and Physics, University of Science and Technology Beijing, Beijing 100083, China}

\author{Zhonghao Xia}
\affiliation{Key Laboratory of Advanced Materials and Devices for Post-Moore Chips, Ministry of Education, University of Science and Technology Beijing, Beijing 100083, China}
\affiliation{School of Mathematics and Physics, University of Science and Technology Beijing, Beijing 100083, China}

\author{Yi Xia}
\affiliation{Department of Mechanical and Materials Engineering, Portland State University, Portland, OR 97201, USA}

\author{Jiangang He}
\email{jghe2024@ustb.edu.cn}
\affiliation{Key Laboratory of Advanced Materials and Devices for Post-Moore Chips, Ministry of Education, University of Science and Technology Beijing, Beijing 100083, China}
\affiliation{School of Mathematics and Physics, University of Science and Technology Beijing, Beijing 100083, China}

\date{\today}


\begin{abstract}
Lattice thermal conductivity ($\kappa_{\rm L}$) is a crucial characteristic of crystalline solids with significant implications for thermal management, energy conversion, and thermal barrier coating. The advancement of computational tools based on density functional theory (DFT) has enabled the effective utilization of phonon quasi-particle-based approaches to unravel the underlying physics of various crystalline systems. While the higher order of anharmonicity is commonly used for explaining extraordinary heat transfer behaviors in crystals, the impact of exchange-correlation (XC) functionals in DFT on describing anharmonicity has been largely overlooked. The XC functional is essential for determining the accuracy of DFT in describing interactions among electrons/ions in solids and molecules. However, most XC functionals in solids focus on ground state properties that mainly involve the harmonic approximation, neglecting temperature effects, and their reliability in studying anharmonic properties remains insufficiently explored. In this study, we systematically investigate the room-temperature $\kappa_{\rm L}$ of 16 binary compounds with rocksalt and zincblende structures using various XC functionals such as local density approximation (LDA), Perdew-Burke-Ernzerhof (PBE), revised PBE for solid and surface (PBEsol), optimized B86b functional (optB86b), revised Tao-Perdew-Staroverov-Scuseria (revTPSS), strongly constrained and appropriately normed functional (SCAN), regularized SCAN (rSCAN) and regularized-restored SCAN (r$^2$SCAN) in combination with different perturbation orders, including phonon within harmonic approximation (HA) plus three-phonon scattering (HA+3ph), phonon calculated using self-consistent phonon theory (SCPH) plus three-phonon scattering (SCPH+3ph), and SCPH phonon plus three- and four-phonon scattering (SCPH+3,4ph). Our results show that the XC functional exhibits strong entanglement with perturbation order and the mean relative absolute error (MRAE) of the computed $\kappa_{\rm L}$ is strongly influenced by both the XC functional and perturbation order, leading to error cancellation or amplification. The minimal (maximal) MRAE is achieved with revTPSS (rSCAN) at the HA+3ph level, SCAN (r$^2$SCAN) at the SCPH+3ph level, and PBEsol (rSCAN) at the SCPH+3,4ph level. Among these functionals, PBEsol exhibits the highest accuracy at the highest perturbation order. The SCAN-related functionals demonstrate moderate accuracy but are suffer from numerical instability and high computational costs. Furthermore, the different impacts of quartic anharmonicity on $\kappa_{\rm L}$ in rocksalt and zincblende structures are identified by all XC functionals, attributed to the distinct lattice anharmonicity in these two structures. These findings serve as a valuable reference for selecting appropriate functionals for describing anharmonic phonons and offer insights into high-order force constant calculations that could facilitate the development of more accurate XC functionals for solid materials.

\end{abstract}

\maketitle

\section{Introduction}
Lattice thermal conductivity is a fundamental characteristic of crystalline solids and holds significant importance in various modern technologies. It plays a crucial role in essential processes such as heat dissipation in integrated circuit chips~\cite{MOORE2014163,2}, direct energy conversion between thermal and electrical energy~\cite{CRC-Handbook,bell2008cooling}, and safeguarding devices using thermal barrier coatings~\cite{doi:10.1021/acs.iecr.1c00788,padture2002thermal,1}. Therefore, the identification of materials with the desired $\kappa_{\rm L}$ values is of significant interest for various technical applications. However, the synthesis and experimental measurement of $\kappa_{\rm L}$ for a specific compound can be a time-consuming and costly process. With the rapid advancements in computer science and simulation theory, many simulation methods have become more efficient in predicting $\kappa_{\rm L}$ of a wide range of crystalline materials. The accuracy of these predictions is highly dependent on the methodology utilized. Different approaches may yield conflicting conclusions~\cite{PhysRevLett.111.025901,PhysRevB.100.245203,RN37}.

Currently, there are primarily two methods used for calculating $\kappa_{\rm L}$ in literature: molecular dynamics (MD), which includes both equilibrium and non-equilibrium MD~\cite{10.1063/1.1740082,Evans_Morriss_2008}, and the Boltzmann's kinetic approach~\cite{KLEMENS19581}. MD accounts for all levels of anharmonic terms and can analyze a wide range of systems such as amorphous materials, nanostructures, and fluids. However, MD simulations require extensive computational time and do not consider quantum effects~\cite{PhysRevMaterials.3.085401}. On the other hand, Boltzmann's kinetic approach offers improved computational efficiency for studying bulk crystals, but its precision depends on the order of perturbation in phonon-phonon interactions~\cite{PhysRevMaterials.3.085401}. For instance, the Peierls-Boltzmann method, which is based on three-phonon interactions, tends to overestimate $\kappa_{\rm L}$ in materials like BAs~\cite{PhysRevLett.111.025901}, with a more accurate value achieved by incorporating four-phonon scattering~\cite{PhysRevB.96.161201}. Similar challenges arise in materials like AgCl and HgTe~\cite{RN37}. However, the computational expense of including higher-order anharmonic effects increases substantially~\cite{HAN2022108179,10.1063/1.5040887}.

Another critical factor that greatly influences the value of the calculated $\kappa_{\rm L}$ is the XC functional utilized in DFT. It is widely recognized that the XC functional is essential in ensuring the precision and reliability of DFT calculations. It is worth noting that various functionals offer distinct representations of the gradual behavior with respect to density gradient, even when operating at the same rung of the Jacob's ladder~\cite{10.1063/1.1390175}. Therefore, it is expected that force constants may be greatly influenced by the chosen XC functional, ultimately impacting $\kappa_{\rm L}$ substantially~\cite{10.1063/5.0173762}. Nevertheless, the development of XC functionals for all properties poses a considerable challenge, as most functionals can only yield precise results for certain properties. With significant dedication to the development of XC functionals over the course of more than 50 years, more than 200 XC functionals have been developed to date. These functionals have been tailored to address a diverse range of properties and scenarios~\cite{200}. Within the solid-state research community, the most commonly utilized XC functionals include LDA~\cite{kohn1965self}, PBE~\cite{PhysRevLett.77.3865}, PBEsol~\cite{PBEsol}, and revTPSS~\cite{PhysRevLett.103.026403}. Recently, the SCAN functional has been developed with meticulous attention to satisfying all 17 known exact constraints~\cite{PhysRevLett.115.036402}. This functional has demonstrated excellent performance across various properties of solids, including lattice constants, polymorph stabilities, phonon dispersion, and more~\cite{PhysRevLett.115.036402,sun2016accurate,PhysRevB.93.045132,PhysRevB.96.035143}. However, subsequent studies have revealed that the SCAN functional exhibits numerical instability and has the potential to lead to divergence in self-consistent calculations in certain instances~\cite{bartok2019regularized,furness2020accurate}. To address this issue, the regularized SCAN (rSCAN) functional was developed, which overcomes the numerical stability problem but at the cost of breaking certain constraints~\cite{bartok2019regularized}. However, the rSCAN functional has been found to exhibit larger errors compared to SCAN in specific systems. To mitigate these errors, the regularized-restored SCAN (r$^2$SCAN) functional was introduced, which achieves a better balance between accuracy and efficiency~\cite{furness2020accurate}. While the performance of these XC functionals has been assessed in our previous work regarding anharmonicity within the quasi-harmonic approximation~\cite{PhysRevB.108.024306}, their performances in computing the $\kappa_{\rm L}$ have not been thoroughly tested yet.

The casual observations in literature have already demonstrated the importance of the XC functional in the calculation of $\kappa_{\rm L}$. For example, Jain and McGaughey explored the influence of XC functionals on the calculated $\kappa_{\rm L}$ for pure crystalline silicon and determined that the relative error in calculated $\kappa_{\rm L}$ can reach up to 17 \%~\cite{RN24}. LDA, PBE, PBEsol, and PW91~\cite{PhysRevB.46.6671} underestimate $\kappa_{\rm L}$, while BLYP~\cite{PhysRevA.38.3098,PhysRevB.37.785} overestimates $\kappa_{\rm L}$ by 12 \%. Qin {\it et al.} investigated the $\kappa_{\rm L}$ of graphene using ten XC functionals and found that the $\kappa_{\rm L}$ values ranged from 1396 to 4376 Wm$^{-1}$K$^{-1}$, depending on the specific XC functional employed~\cite{RN24}. Taheri {\it et al.} achieved higher values of $\kappa_{\rm L}$ for graphene, ranging from 5442 to 8677 Wm$^{-1}$K$^{-1}$ by utilizing five different XC functionals and two types of pseudopotentials~\cite{RN25,RN26}. Recently, Han and Ruan discovered that four-phonon scattering plays a significant role in graphene, and the discrepancies found in the literature regarding the calculated $\kappa_{\rm L}$ are attributed to convergence issues in the grid sampling of the Brillouin zone and the presence of higher orders of anharmonicity~\cite{PhysRevB.108.L121412}. In a separate study, Dongre {\it et al.} determined that the calculated $\kappa_{\rm L}$ of GaP using PBE is 1.8 times larger than that calculated value using LDA~\cite{RN28}. Xia {\it et al.} investigated the $\kappa_{\rm L}$ of Tl$_3$VSe$_4$ using the PBEsol functional~\cite{xia2020particlelike}, while Jain conducted similar calculations for this compound using PBE~\cite{jain2020multichannel}. The discrepancies in the calculated $\kappa_{\rm L}$ values obtained using these two XC functionals have led to different explanations regarding the origin of the ultralow $\kappa_{\rm L}$ in this compound~\cite{xia2020particlelike,jain2020multichannel}. It is worth noting that the interplay between the XC functional and the perturbation order of anharmonicity complicates the underlying physics of $\kappa_{\rm L}$, as errors introduced by the XC functional and anharmonicity order may inadvertently cancel each other out, resulting in better agreement with experimental values when using low-order anharmonicity approaches, as demonstrated in previous studies~\cite{RN37}.

In this study, we systematically evaluate the performance of three commonly used functionals (LDA, PBE, and PBEsol) in solids, one popular van der Waals density functional (optB86b)~\cite{PhysRevB.83.195131}, and four meta-GGA functionals (revTPSS, SCAN, rSCAN, and r$^2$SCAN) in predicting the room-temperature $\kappa_{\rm L}$ for 16 binary compounds with rocksalt and zincblende structures at three perturbation orders, namely HA+3ph, SCPH+3ph, and SCPH+3,4ph. Our results show that the strong entanglement between the XC functional and the perturbation order complicates the comparison between the computed and the experimental $\kappa_{\rm L}$. In other words, the MRAE of the computed $\kappa_{\rm L}$ is strongly influenced by both the XC functional and the perturbation order, leading to error cancellation or amplification. The minimal (maximal) MRAE is achieved with revTPSS (rSCAN) at the HA+3ph level, SCAN (r$^2$SCAN) at the SCPH+3ph level, and PBEsol (revTPSS) at the SCPH+3,4ph level. Overall, PBEsol is numerically stable and exhibits the highest accuracy in calculating $\kappa_{\rm L}$ among these functionals. The SCAN-related functionals demonstrate moderate accuracy but suffer from numerical instability and higher computational cost. Furthermore, all XC functionals identify the different impacts of quartic anharmonicity on $\kappa_{\rm L}$ between rocksalt and zincblende structures, which can be attributed to their distinct coordination environment and lattice anharmonicity. These findings serve as a valuable reference for selecting appropriate functionals for describing anharmonic phonons and offer insights into high-order force constant calculations that could facilitate the development of more accurate XC functionals for solid materials.

\section{COMPUTATION  DETAILS}
All DFT calculations were performed using the projector-augmented wave (PAW)\cite{PAW1,PAW2} method within the Vienna {\it Ab initio} Simulation Package (VASP)~\cite{vasp1,vasp2}. Eight XC functionals were utilized, namely LDA, PBE, PBEsol, optB86b, revTPSS, SCAN, rSCAN, and r$^2$SCAN. With the exception of LDA, the PBE-version PAW pseudopotential was employed for all other functionals due to the unavailability of the PAW pseudopotentials for these functionals in VASP. While this may introduce some errors in these functionals, previous studies on solids have predominantly used the PBE pseudopotential, with negligible errors reported in studies utilizing the other functional~\cite{RN52,RN53,RN54}. The plane-wave basis set was used with cutoff energies of 520 eV for LDA, PBE, PBEsol, and optB86b, 600 eV for revTPSS, and 800 eV for SCAN, rSCAN, and r$^2$SCAN. Additionally, for SCAN, rSCAN, and r$^2$SCAN, the precision parameter was set to "accurate" (PREC = ACCURATE), nonspherical contributions within the PAW spheres were included (LASPH = .TRUE.), and additional grids for the evaluation of augmentation charges (ADDGRID = .TRUE.) were implemented.

The second-order force constants were calculated using the finite displacement method as implemented in Phonopy code~\cite{TOGO20151} with 4 $\times$ 4 $\times$ 4 supercell and 2 $\times$ 2 $\times$ 2 $k$-points mesh, and a 0.01 \AA displacement. During the self-consistent calculations, the energy is converged to $10^{-8}$ eV. The projection operators were evaluated in reciprocal space (LREAL = .FALSE.) to more accurately compute forces. Since all atoms are in high-symmetry Wyckoff positions and the only structure parameter is the lattice constant, the equilibrium structures of all compounds studied in this work were generated using the experimental lattice constants at room temperature to eliminate errors caused by thermal expansion. The effect of supercell size on harmonic phonons was carefully tested. It was found that a 5 $\times$ 5 $\times$ 5 supercell can produce artificial imaginary frequencies for some compounds using certain XC functionals, while the more numerically stable option for all compounds was the 4 $\times$ 4 $\times$ 4 supercell, as shown in Fig.~\textcolor{magenta}{S1}. The nonanalytic correction to phonon dispersions near the $\Gamma$ point is performed by dipole-dipole interaction~\cite{PhysRevB.50.13035,PhysRevB.55.10355}, based on the PBEsol functional calculated Born effective charges and macroscopic static dielectric constants using the density function perturbation theory implementation in VASP.

The third-order and fourth-order force constants were extracted using the compressive sensing lattice dynamics (CSLD)~\cite{RN38}. Phonon frequency shifts at finite temperatures were calculated using the SCPH theory~\cite{doi:10.1080/14786440408520575,PhysRevLett.17.753,PhysRevB.1.572}. The Peierls-Boltzmann transport equation was literately solved by sampling with a uniform 24 $\times$ 24 $\times$ 24 $q$-point mesh for $\kappa_{\rm L}$ calculations with HA+3ph ($\kappa_{\rm 3ph}^{\rm HA}$) and SCPH+3ph ($\kappa_{\rm 3ph}^{\rm SCPH}$) methods, as well as a sparser 16 $\times$ 16 $\times$ 16 $q$-point mesh for $\kappa_{\rm L}$ with SCPH+3,4ph ($\kappa_{\rm 3,4ph}^{\rm SCPH}$) method. The four-phonon scattering process is accelerated using the sampling method~\cite{guo2024sampling}. All other parameters remain consistent with the previous study~\cite{RN37}.

Elastic constants are determined through the structure deformation method as implemented in Pymatgen~\cite{ONG2013314}. The mean sound velocity ($v_{\rm m}$) is obtained as the average of the longitudinal ($v_{\rm L}$) and transverse ($v_{\rm T}$) sound velocities, which are calculated based on the density ($\rho$), bulk ($B$), and shear ($G$) modulus~\cite{R-Hill_1952}.

$v_m=[\frac{1}{3}(\frac{1}{v_{\rm L}^3}+\frac{2}{v_{\rm T}^3})]^{\frac{-1}{3}}$

$v_{\rm T} = \sqrt{\frac{G}{\rho}}$

$v_{\rm L} = \sqrt{\frac{B+\frac{4}{3}G}{\rho}}$

\section{RESULTS AND DISCUSSION}
In the kinetic theory~\cite{tritt2005thermal}, $\kappa_{\rm L}$ can be represented as the product of the heat capacity ($C_{\rm v}$), the square of the phonon group velocity ($v_{\rm g}$), and the phonon relaxation time ($\tau$), as shown in the equation $\kappa_{\rm L} = \frac{1}{3}C_{\rm v}v_{\rm g}^2\tau$. Therefore, prior to assessing the accuracy of computing $\kappa_{\rm L}$ using eight XC functionals at three theoretical perturbation orders, a comparison was conducted on the precision of these functionals regarding phonon spectra, $v_{\rm g}$, and the Gr\"uneisen parameter ($\gamma$). The Gr\"uneisen parameter is commonly utilized to characterize anharmonicity and is closely linked to the scattering rate ($1/\tau$).

\subsection{Phonon spectrum}
\begin{figure}[tph!]
	\includegraphics[clip,width=1.0\linewidth]{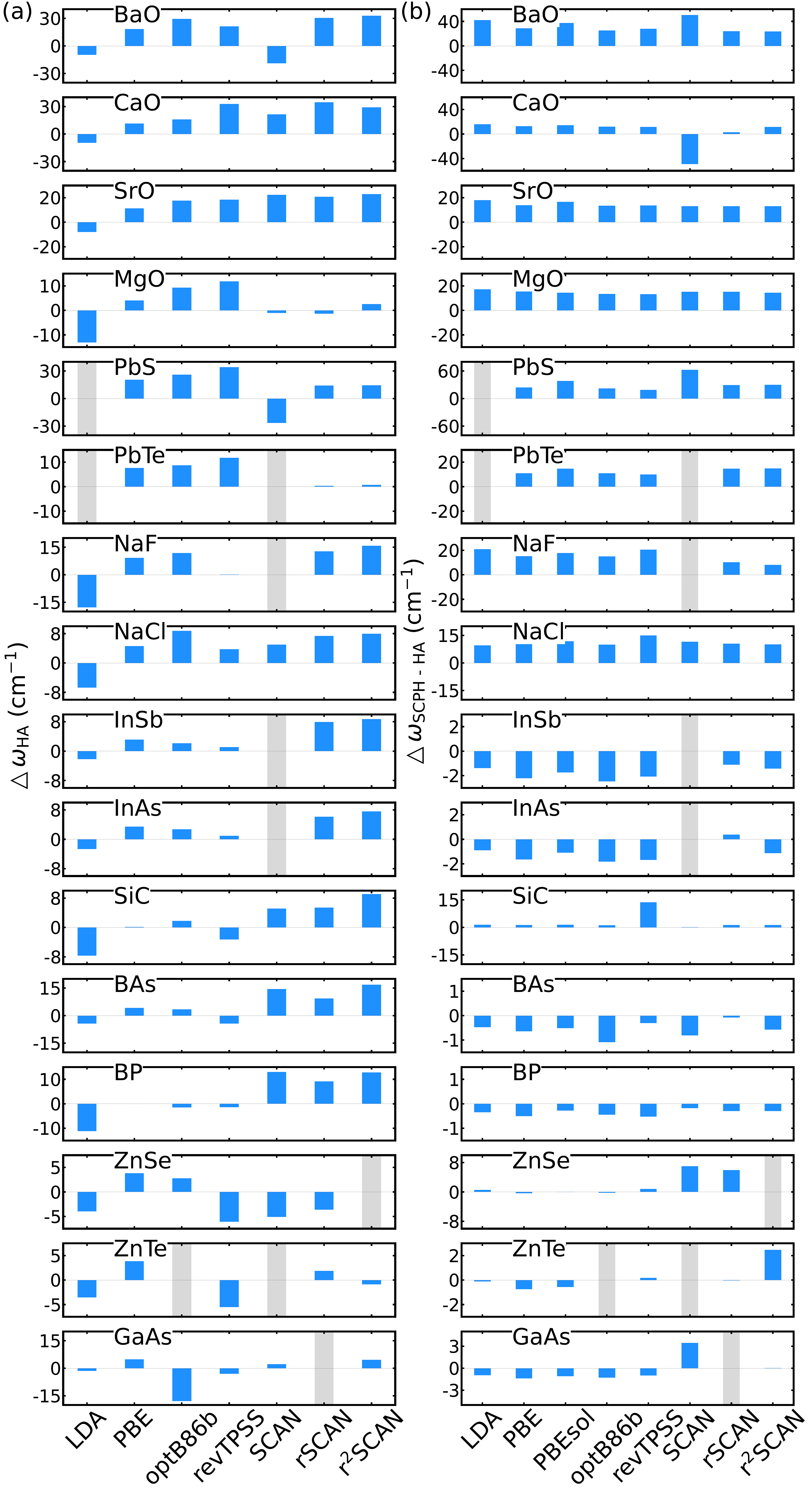}
	\caption{(Color online) (a) The HA frequency difference ($\Delta{\omega}_{\rm HA}$) of the optical phonon frequency at the $\Gamma$ point calculated using the other functionals with respect to that obtained with PBEsol. (b) The frequency shift ($\Delta{\omega}_{\rm SCPH - HA}$) between the phonon calculated by SCPH and HA for all the compounds calculated with 8 XC functionals.}
	\label{fig:phonon}
\end{figure}

The phonon spectrum contains essential information regarding lattice dynamics, rendering it indispensable for comprehending $\kappa_{\rm L}$, and is also significantly influenced by temperature. However, the harmonic phonon, obtained from the harmonic approximation by diagonalizing the dynamical matrix formed from harmonic interatomic force constants, fails to consider temperature effects. The impact of temperature on the phonon spectrum can be effectively simulated using the SCPH theory~\cite{doi:10.1080/14786440408520575,PhysRevLett.17.89,PhysRevB.1.572}. In this study, the phonon spectrum at room temperature is calculated using the first-order correction based on quartic anharmonicity~\cite{PhysRevLett.125.085901}. The phonon spectra for all the compounds are analyzed using both harmonic and quartic anharmonic approximations, with the results computed using 8 XC functionals presented in Fig.~\textcolor{magenta}{S2-S7}. Additionally, Fig.~\ref{fig:phonon}(a) illustrates the discrepancy ($\Delta{\omega}_{\rm HA}$) in the optical phonon frequency at the $\Gamma$ point when calculated using different XC functionals compared to the PBEsol functional. Note the phonon frequency before nonanalytical correction is adopted here for the sake of simplicity. It is unexpected that the $\Delta{\omega}_{\rm HA}$ of the rocksalt compounds is greater than that of the zincblende, even though the rocksalt compounds have a considerably lower optical frequency at the $\Gamma$ point, indicating that the phonon frequencies of the zincblende compounds are less sensitive to the XC functional than rocksalt compounds. Out of all these functionals, only LDA exhibits negative $\Delta{\omega}_{\rm HA}$ for all the compounds due to the fact that LDA calculated chemical bonds are elongated (bonding interaction is weakened) when the experimental volumes utilized in this study. When comparing the phonon frequencies calculated using the HA with those obtained through the SCPH method at 300 K, it is observed that the rocksalt structure exhibits a slight increase in frequency, particularly in the case of three optical branches. Conversely, the phonon frequency of the zincblende structure remains relatively unchanged or even decreases slightly in certain instances. In order to quantify this, we show the frequency shift of the phonon at the $\Gamma$ point from the SCPH to HA method ($\Delta{\omega}_{\rm SCPH - HA}$ = $\omega_{\rm SCPH}$ - $\omega_{\rm HA}$) in Fig.~\ref{fig:phonon}(b). All the $\Delta{\omega}_{\rm SCPH - HA}$ of the rocksalt compounds are large and positive except that of CaO calculated by SCAN, while the $\Delta{\omega}_{\rm SCPH - HA}$ of the zincblende compounds are generally small or even negative by some of these functionals. Exceptions are predominantly observed in four metaGGA functionals: InAs by rSCAN, SiC by revTPSS, rSCAN, and r$^2$SCAN, ZnSe by revTPSS, SCAN, and rSCAN, ZnTe by r$^2$SCAN, and GaAs by SCAN. This highlights the differences of the role played by quartic anharmonicity between rocksalt and zincblende structures. All functionals predict a large positive $\Delta{\omega}_{\rm SCPH - HA}$ in BaO, while revTPSS is the only functional predicting a large $\Delta{\omega}_{\rm SCPH - HA}$ in SiC and SCAN is the only functional having a large and negative $\Delta{\omega}_{\rm SCPH - HA}$ for CaO. These observations underscore the large difference between metaGGA and semilocal functionals in characterizing anharmonicity.

\subsection{Sound velocity}
\begin{figure}[tph!]
	\includegraphics[clip,width=1.0\linewidth]{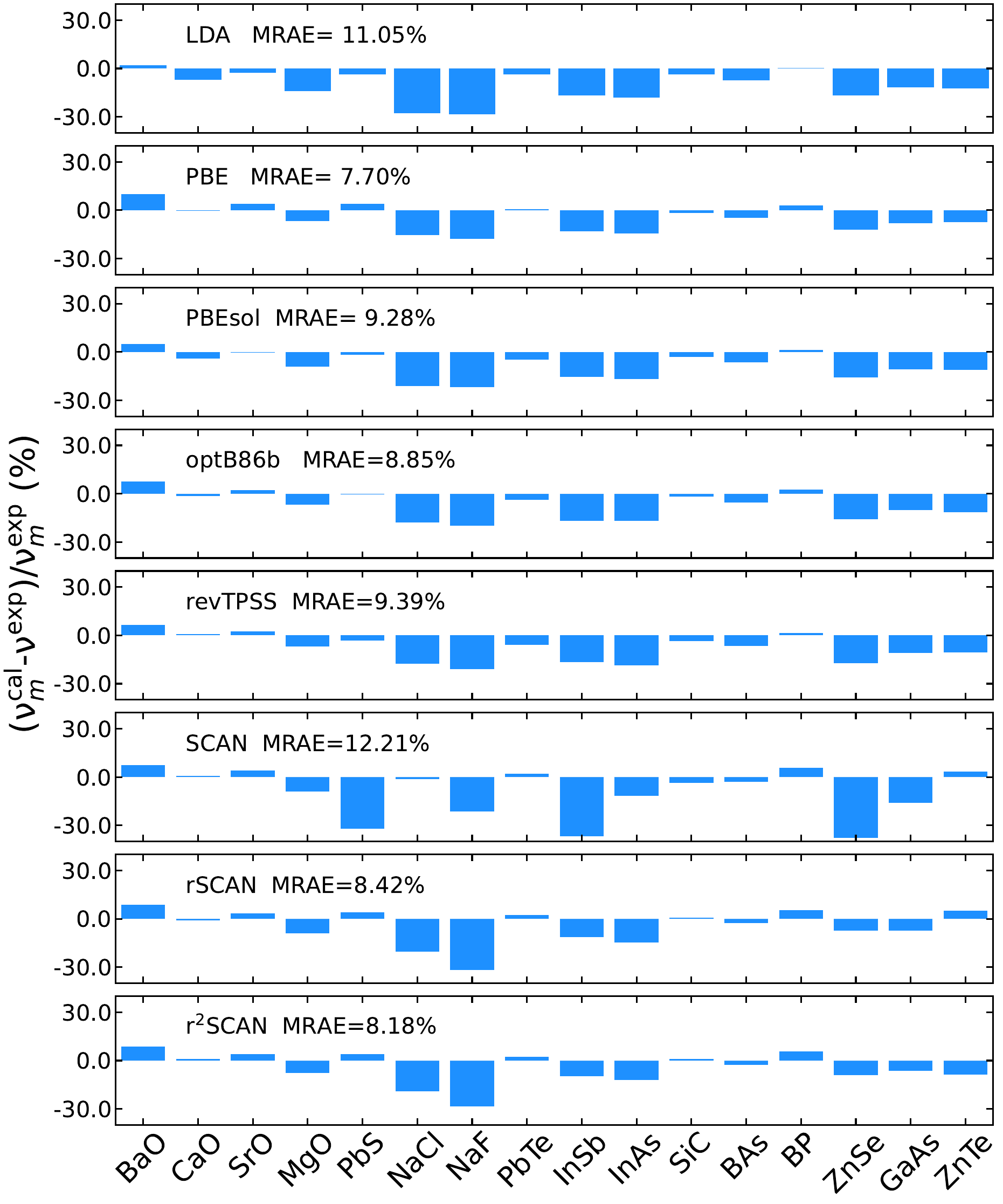}
	\caption{(Color online). The relative error of the calculated mean sound velocity ($v_{\rm m}^{\rm cal.}$) with respect to experiment values ($v_{\rm m}^{\rm exp.}$). The $v_{\rm m}^{\rm exp.}$ are either taken from Ref.~\onlinecite{chen2018rationalizing} or calculated using the experimental elastic constants in Ref.~\onlinecite{chen2018rationalizing}.}
	\label{fig:groupvelocity}
\end{figure}

Phonon group velocity $v_{\rm g}$ is another pivotal metric encapsulating the speed at which the main heat carrying phonons propagate through a solid crystalline, determining its thermal transport behavior~\cite{https://doi.org/10.1002/adma.201900108}. Therefore, it is widely used in both empirical formula and phonon Boltzmann transport theory for computing $\kappa_{\rm L}$. Herein, $v_{\rm g}$ is approximated by the mean sound velocity ($v_{\rm m}$), which is calculated from elastic constants, see the ``Computational Details'' section for details. The $v_{\rm m}$ is a critical parameter that measures the speed of the primary heat-carrying phonons traveling through a solid crystalline material, influencing its thermal conductivity behavior~\cite{https://doi.org/10.1002/adma.201900108}. Figure~\ref{fig:groupvelocity} illustrates the comparison between our calculated $v_{\rm m}$ values of all the compounds with the experimental values measured at room temperature, utilizing all the XC functionals examined in this study. It should be noted that the majority of $v_{\rm m}$ values are derived from direct experimental measurements, while a small subset are calculated using the experimental elastic constants. The calculated $v_{\rm m}$ values of nearly all compounds, as determined using all the XC functionals, are consistently lower than the corresponding experimental values, with the notable exception of BaO. This discrepancy is unexpected, considering that our calculations are focused solely on the intrinsic compounds and do not account for potential factors such as defects and grain boundaries, which are known to exert a substantial slowing down of $v_{\rm m}$~\cite{https://doi.org/10.1002/adma.201900108}. The MRAE of these compounds exhibit only slight variations across the XC functionals. The smallest relative errors are observed at CaO and SiC, while the largest errors are found at NaF and NaCl. Specifically, the smallest and largest MRAE values are observed in PBE (7.7 \%) and SCAN (12.2 \%) functionals, respectively. It is noteworthy that the calculated $v_{\rm m}$ values for NaCl and NaF show significant deviations from the experimental values. SCAN, among all the XC functionals, identifies PbS, ZnSe, and InSb as having the largest relative errors, leading to the highest MRAE, despite its smallest relative error in NaCl. In general, the variation in $v_{\rm m}$ across various functionals is relatively small presumably due to the fact that the elastic constant is the second derivative of energy with respect to displacement, which is harmonic and can be well described by all these functionals.

\subsection{Gr\"uneisen parameter}
\begin{figure*}[tph!]
	\includegraphics[clip,width=0.8\linewidth]{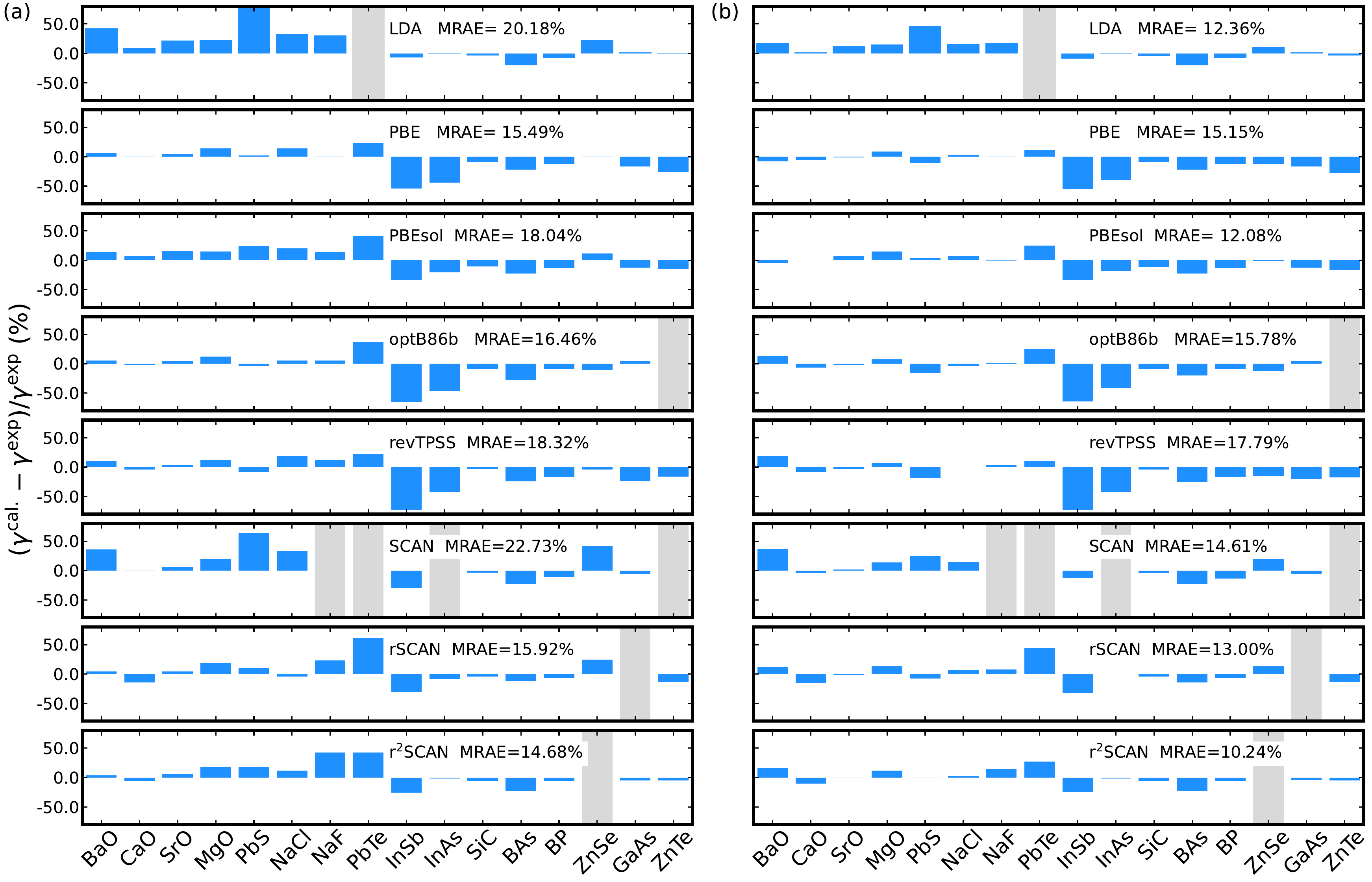}
	\caption{(Color online). (a) and (b) are the relative errors of the total Gr\"uneisen parameters ($\gamma$) at 0 K and 300 K, respectively, when compared to experimental values. Gray color represents the calculation with imaginary frequency at 0 K (a) and at room temperature (b).}
	\label{fig:gruneisen}
\end{figure*}

The Gr\"uneisen parameter $\gamma$, which is defined as the rate of change of phonon frequency in response to variations in crystal volume~\cite{https://doi.org/10.1002/andp.19083310611,https://doi.org/10.1002/andp.19123441202}, is commonly utilized to quantify the degree of anharmonicity and can be employed to compute $\kappa_{\rm L}$ and the thermal expansion coefficient of solids~\cite{morelli2006high,morelli2006high,ritz2019thermal}. In our previous study~\cite{PhysRevB.108.024306}, we evaluated the performance of six XC functionals on computing the average value of $\gamma$ for 17 binary compounds with rocksalt, zincblende, and fluorite structures. The average $\gamma$ was determined using the Phonopy code~\cite{TOGO20151} based on the HA phonon frequencies of 11 volumes around the equilibrium volume at 0 K. In this study, the total $\gamma$ is computed as a weighted sum of the mode contributions, as implemented in the ShengBTE package~\cite{ShengBTE_2014}, incorporating both the HA and SCPH force constants. Consequently, two $\gamma$ values are determined for each compound, specifically at 0 and 300 K. Figure~\ref{fig:gruneisen}(a) illustrates the comparison between the experimental and calculated $\gamma$ values for all the rocksalt and zincblende compounds examined in this work. The experimental $\gamma$ values primarily originate from Ref.~\onlinecite{shinde2006high,slack1979thermal}. For the calculated $\gamma$ values based on the HA force constants, the MRAEs of the LDA, PBE, PBEsol, optB86b, revTPSS, SCAN, rSCAN, and r$^2$SCAN functionals are 20.18 \%, 15.49 \%, 18.04 \%, 16.46 \%, 18.32 \%, 22.73 \%, 15.92 \%, and 14.68 \%, respectively. It should be noted that both SCAN and r$^2$SCAN exhibit significant discrepancies compared to the previous results~\cite{PhysRevB.108.024306}, which can be attributed to differences in the dataset as well as the impact of lattice constants, \textit{i.e.,} the use of DFT-optimized versus experimental lattice constants. Notably, the compounds that resulted in large relative errors with the SCAN and r$^2$SCAN functionals in our previous work (AlP and InP) were excluded from this study due to the presence of imaginary frequencies when employing the experimental lattice constants. Additionally, the substantial difference observed in InSb may be attributed to the discrepancy in the employed lattice constants. As depicted in Fig.~\ref{fig:gruneisen}(b), the replacement of the HA force constant with the SCPH one leads to a reduction in all MRAEs, particularly for the LDA, PBEsol, and SCAN functionals, primarily due to improved agreement for PbS. Among all the XC functionals considered, r$^2$SCAN exhibits the lowest MRAE (10.24 \%), followed by PBEsol (12.08 \%) and LDA (12.36 \%). With the exception of LDA and rSCAN, all functionals significantly underestimate the $\gamma$ value of InSb. It is evident that PBE, PBEsol, optB86b, and revTPSS functionals yield smaller MRAEs for the $\gamma$ values of rocksalt compounds compared to zincblende ones, likely due to the generally larger $\gamma$ values observed in rocksalt structure than zincblende~\cite{doi:10.1021/acs.chemmater.6b04179}.

\subsection{Scattering rates}
The main process of phonon scattering in an intrinsic semiconductor is the phonon-phonon scattering, predominantly involving a three-phonon process~\cite{peierls1996quantum}. However, four-phonon scattering may also be significant in certain systems~\cite{PhysRevB.96.161201,PhysRevLett.125.245901}. The scattering rate (1/$\tau$) characterizes the strength of scattering among phonons and is a key factor in the phonon Boltzmann transport equation, playing a dominant role in determining $\kappa_{\rm L}$~\cite{PhysRevB.93.045202}. In this study, scattering rates associated with the three-phonon and four-phonon processes are calculated using the single-mode relaxation approximation~\cite{PhysRevB.93.045202}. As depicted in Fig.~\textcolor{magenta}{S8-S13}, all XC functionals indicate that three-phonon scattering prevails in the scattering process across all compounds. Specifically, the three-phonon scattering rates in the low-frequency region, a key factor in heat resistance, are higher in rocksalt compounds ($\sim$ 10 ps$^{-1}$) compared to those with the zincblende structure ($\sim$ 1 ps$^{-1}$), consistent with previous research and reflecting the stronger anharmonicity (larger $\gamma$) in rocksalt structure compared to zincblende~\cite{RN37,https://doi.org/10.1002/anie.201508381}. Furthermore, all XC functionals suggest that the four-phonon scattering rates in zincblende compounds are relatively higher than those in rocksalt compounds, aligning with previous observations based on PBE functional~\cite{RN37}, due to the relatively smaller three-phonon scattering rates in zincblende compounds.

By comparing the phonon scattering rates of a compound computed using different XC functionals, it is evident that SCAN-related functionals have a higher likelihood of predicting a different scattering rate from the other XC functionals, as illustrated in Fig.~\textcolor{magenta}{S9-S11}. For example, rSCAN and r$^2$SCAN consistently show significantly lower three-phonon scattering rates for NaCl and NaF compared to other XC functionals. SCAN predicts notably higher three-phonon scattering rates for PbS and SiC in the low-frequency region. rSCAN demonstrates significantly higher three-phonon scattering rates for NaCl, NaF, and BAs in comparison to other functionals, while r$^2$SCAN predicts a notably high three-phonon scattering rate for ZnTe in the low-frequency region. Furthermore, rSCAN and r$^2$SCAN predict exceptionally high four-phonon scattering rates for NaF in the frequency range of 150 $\sim$ 250 cm$^{-1}$, with rSCAN also showing a remarkably high four-phonon scattering rate for NaCl. These findings underscore the complexity and challenges associated with computing three- and four-phonon scattering rates, especially for metaGGA functionals.

\subsection{Lattice thermal conductivity at room temperature}
The room-temperature $\kappa_{\rm L}$ of 16 binary compounds with rocksalt and zincblende structures are computed using 8 commonly used XC functionals in solids at three different perturbation orders: HA+3ph, SCPH+3ph, and SCPH+3,4ph. Currently, the mainstream method of computing $\kappa_{\rm L}$ is HA+3ph, particularly for complex systems. SCPH+3ph accounts for phonon frequency shifts due to temperature effect, necessitating the calculation of higher-order force constants, and is particularly relevant for systems with multiple low-frequencies phonon branches. SCPH+3,4ph is considered as the most accurate method in this study, as it incorporates both frequency shifts due to finite temperature and four-phonon scattering.

\begin{figure*}[tph!]
	\includegraphics[clip,width=1.0\linewidth]{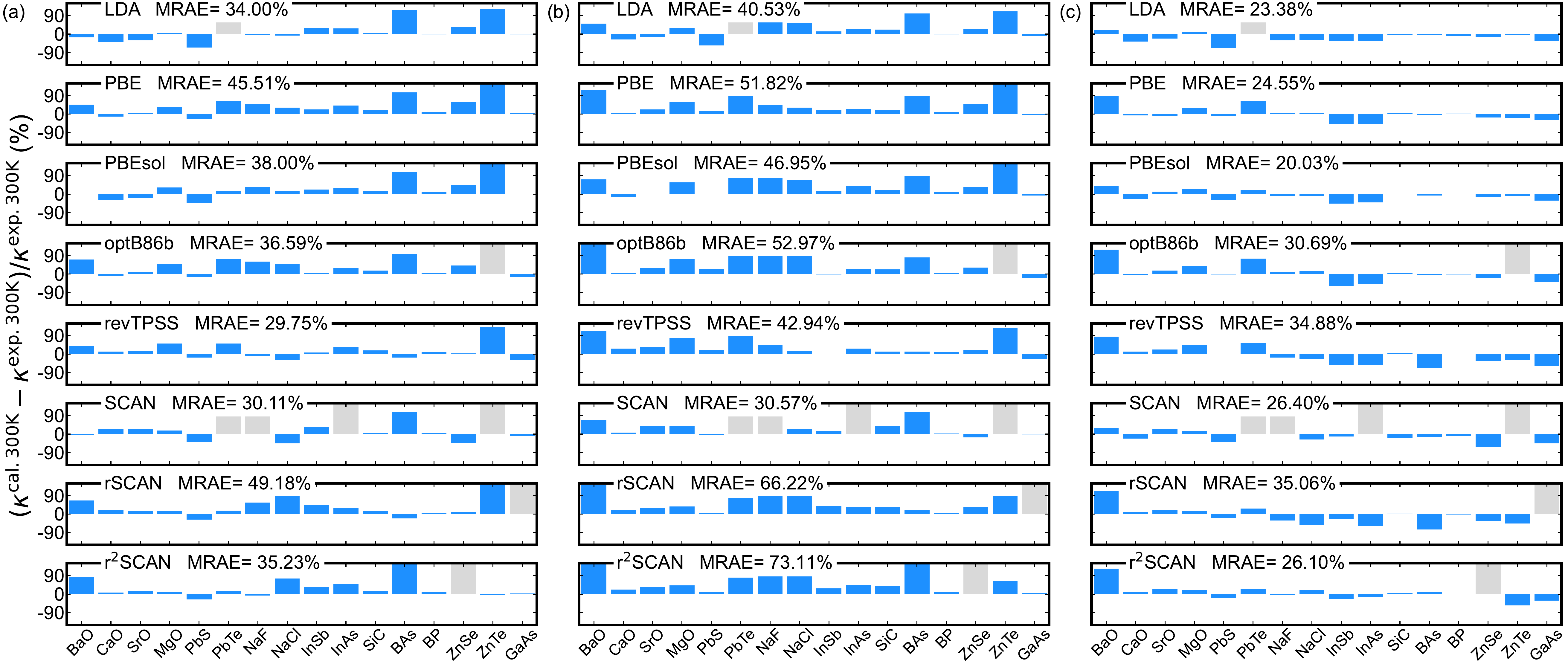}
	\caption{(Color online). (a)-(c) are relative errors of the calculated $\kappa_{\rm L}$ at HA+3ph ($\kappa_{\rm 3ph}^{\rm HA}$), SCPH+3ph ($\kappa_{\rm 3ph}^{\rm SCPH}$), and SCPH+3,4ph ($\kappa_{\rm 3,4ph}^{\rm SCPH}$) three levels with respect to the experimental values at room temperature, respectively. Gray color indicates the phonon with imaginary frequencies at 0 K or 300 K that prohibits $\kappa_{\rm L}$ calculations. Experimental $\kappa_{\rm L}$ values are adopted from~\onlinecite{simoncelli2019unified,el1983thermophysical,akhmedova2009effect,haakansson1985thermal,haakansson1986thermal,steigmeier1963thermal,madelung2004semiconductors,kang2018experimental,kumashiro1989thermal,slack1972thermal}.}
	\label{fig:ha+scph}
\end{figure*}

The relative error of the calculated $\kappa_{\rm L}$ for 16 compounds at the HA+3ph level with 8 XC functionals is depicted in Fig.~\ref{fig:ha+scph}(a). It is important to note that some compounds exhibit significant imaginary frequencies either at the experimentally determined room-temperature lattice constants or due to artificial numerical issues associated with specific functionals, as previously observed in our lattice constant study~\cite{PhysRevB.108.024306}. Some of these imaginary frequencies cannot be stabilized at room temperature using the SCPH method either. The presence of imaginary frequencies at 0 K prevents the calculation of $\kappa_{\rm 3ph}^{\rm HA}$, while those at room temperature impede the determination of $\kappa_{\rm 3ph}^{\rm SCPH}$ and $\kappa_{\rm 3,4ph}^{\rm SCPH}$. The minimum and maximum MRAEs of the calculated $\kappa_{\rm L}$ at the HA+3ph level are 29.75 \% using revTPSS and 49.18 \% using rSCAN, primarily attributed to the overestimation of $\kappa_{\rm L}$ in BAs, which is a compound known for strong four-phonon scattering~\cite{PhysRevB.100.245203}. Among all the functionals considered in this work, only LDA and SCAN show significant underestimation of $\kappa_{\rm L}$ for compounds with rocksalt structure. Additionally, PBE underestimates $\kappa_{\rm L}$ only for CaO and PbS, which is contrary to the previous finding that PBE underestimates $\kappa_{\rm 3ph}^{\rm HA}$ for most rocksalt compounds~\cite{RN37}. This discrepancy arises from the use of experimentally determined room-temperature lattice constants in this study, whereas the previous work utilized the PBE optimized lattice constants at 0 K~\cite{RN37}.

\begin{figure}[tph!]
	\includegraphics[clip,width=1.0\linewidth]{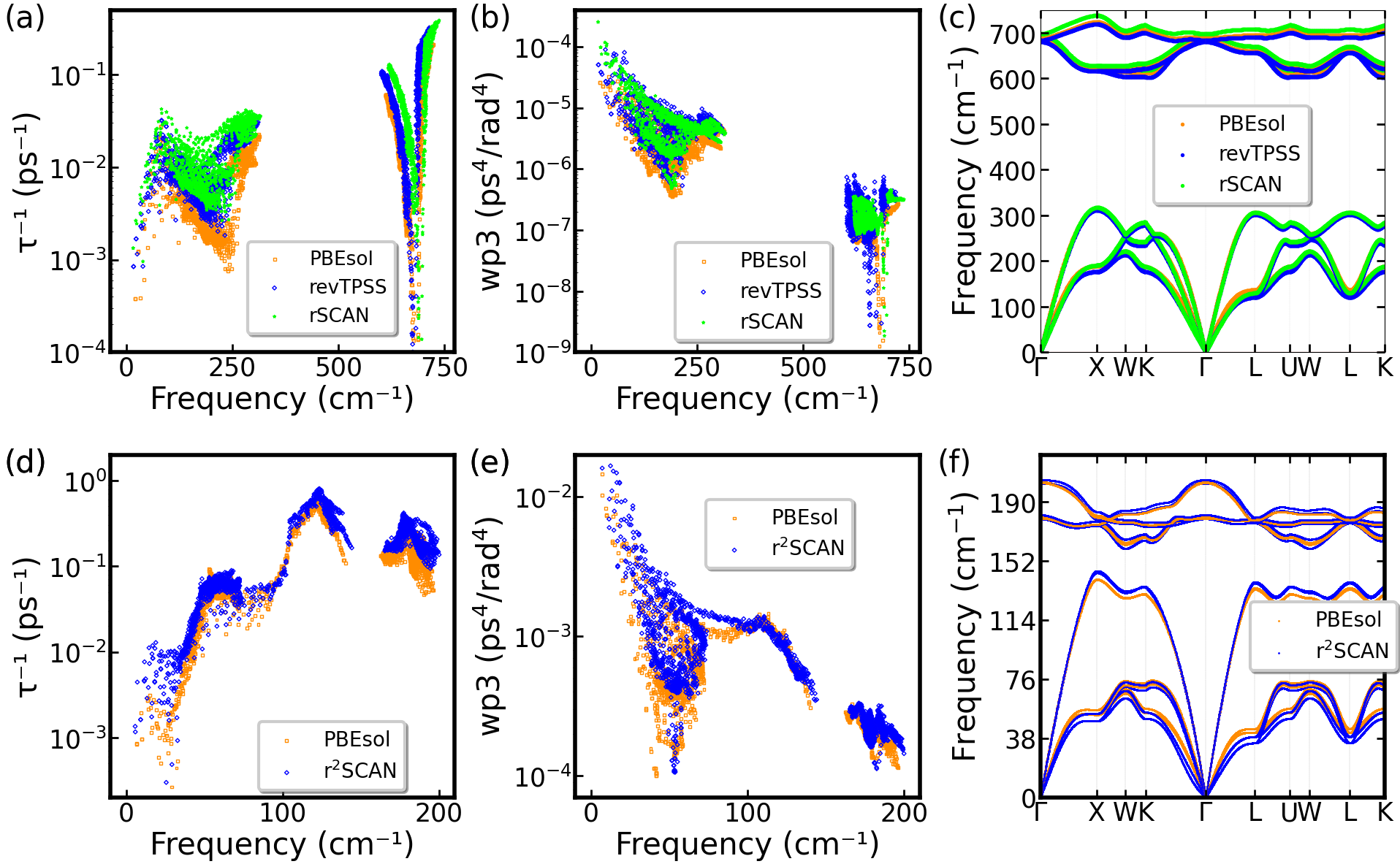}
	\caption{(Color online) (a)-(c) are the three-phonon scattering rate, weighted phase space, and phonon spectrum of BAs, respectively. (d)-(f) are the three-phonon scattering rate, weighted phase space, and phonon spectrum of ZnTe, respectively.}
	\label{fig:BAs-ZnTe}
\end{figure}

As illustrated in Fig.~\textcolor{magenta}{S14}, the experimental volumes measured at room temperature are smaller than the PBE optimized ones at 0 K for all compounds, resulting in shorter bond lengths and stronger bond strength, leading to enlarged $\kappa_{\rm L}$ values. OptB86b, similarly, overestimates the $\kappa_{\rm L}$ for all compounds except PbS. However, this trend varies among different functionals. For instance, while revTPSS exhibits larger optimized volumes compared to optB86b (see Fig.~\textcolor{magenta}{S14}), the overestimation of $\kappa_{\rm L}$ by optB86b surpasses that of revTPSS. The majority of XC functionals analyzed in this study have a tendency to overestimate the $\kappa_{\rm L}$ values for compounds with zincblende structure, with BAs and ZnTe exhibiting the most significant overestimations. This is consistent with the results reported by Xia \textit{et al.} utilizing the PBE functional with optimized lattice constants~\cite{RN37}. Nonetheless, there are exceptions such as ZnTe with r$^2$SCAN, BAs with rSCAN and revTPSS. This is particularly surprising as it is found that the calculated $\kappa_{\rm 3ph}^{\rm HA}$ is typically twice as large as the experimentally measured $\kappa_{\rm L}$ due to its relatively strong four-phonon scattering~\cite{PhysRevB.96.161201,doi:10.1126/science.aaz6149}. Furthermore, as depicted in Fig.~\ref{fig:BAs-ZnTe} and ~\ref{fig:groupvelocity}, the revTPSS and rSCAN functionals exhibit similar phonon spectra and $v_{\rm m}$ to other functionals, such as PBEsol, but display a stronger three-phonon scattering rate at low-frequency regions due to their larger weighted phase space~\cite{PhysRevB.89.184304}. Interestingly, despite being derived from revTPSS~\cite{PhysRevLett.115.036402}, SCAN and r$^2$SCAN do not underestimate the three-phonon scattering rate of BAs and exhibit very similar MRAE. This suggests the importance of adhering to the constraints imposed by SCAN and r$^2$SCAN. Among these functionals, revTPSS functional has the smallest relative error ($\kappa_{\rm 3ph}^{\rm HA}$) for all compounds except ZnTe, resulting in the lowest MRAE (29.75 \%). Although r$^2$SCAN exhibits a similar $v_{\rm m}$ as PBEsol for ZnTe, as shown in Fig.~\ref{fig:groupvelocity}, it has much stronger three-phonon scattering in the low-frequency region, see Fig.~\ref{fig:BAs-ZnTe}, resulting in good agreement with the experimental $\kappa_{\rm L}$. It is worth noting that this agreement is likely a result of error cancellation, as the phonon frequency shifts and scattering rates caused by quadratic anharmonicity have significant effects on $\kappa_{\rm L}$, leading to fortuitous agreement between theory and experiment.

\begin{figure*}[tph!]
	\includegraphics[clip,width=1.0\linewidth]{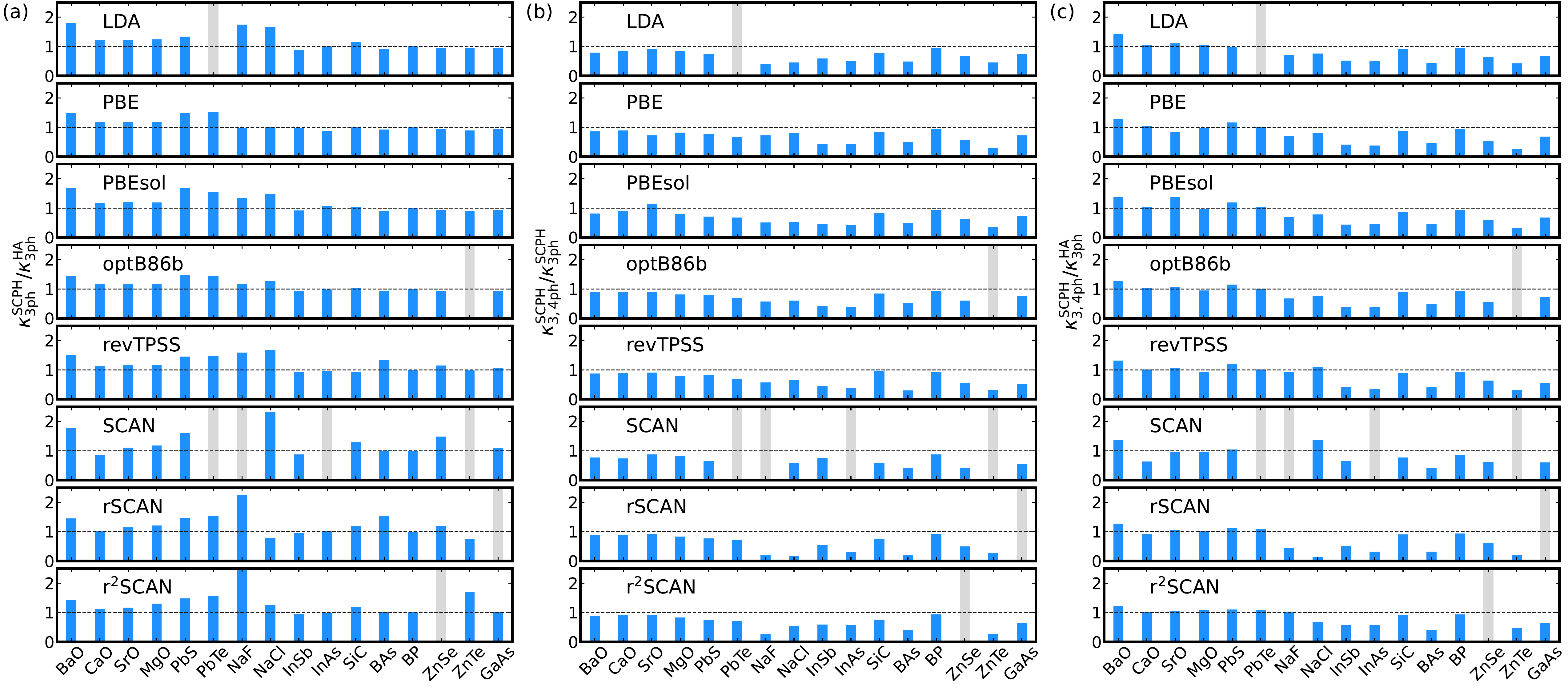}
	\caption{(Color online). Comparison of $\kappa_{\rm L}$ calculated using three methods for the 16 binary with rocksalts and zincblende structure at 300 K. (a)-(c) are the $\kappa_{\rm 3ph}^{\rm SCPH}/\kappa_{\rm 3ph}^{\rm HA}$, $\kappa_{\rm 3,4ph}^{\rm SCPH}/\kappa_{\rm 3ph}^{\rm SCPH}$, and $\kappa_{\rm 3,4ph}^{\rm SCPH}/\kappa_{\rm 3ph}^{\rm HA}$, respectively. Gray color represents the calculation that has imaginary frequency at 0 K or 300 K, and hence the $\kappa_{\rm L}$ calculation is prohibited.}
	\label{fig:ratio}
\end{figure*}

The Fig.~\ref{fig:ha+scph}(b) displays the calculated $\kappa_{\rm L}$ values by eight XC functionals at the SCPH+3ph level. Among these functionals, the SCAN-related functionals yield both the highest and lowest MRAEs, with SCAN exhibiting the smallest MRAE of 30.57 \% and r$^2$SCAN exhibiting the largest MRAE of 73.11 \%. Following closely behind is rSCAN with a MRAE of 66.22 \%. It is evident that all XC functionals show a significant increase in MRAE compared to the $\kappa_{\rm L}$ values calculated using HA+3ph. The largest increase occurs in rSCAN and r$^2$SCAN, with their MRAEs increasing from 49.18 \% to 66.22 \% and from 35.23 \% to 73.11 \%, respectively. The relative errors of each compound obtained from rSCAN and r$^2$SCAN are notably similar, with the exception of BAs, which can be attributed to the strong three-phonon scattering predicted by rSCAN, see Fig.~\ref{fig:BAs-ZnTe}. The increase in MRAE from HA+3ph to SCPH+3ph is mainly attributed to the shift of the phonon frequency towards higher energy (see Fig.~\ref{fig:phonon}) and the prior-existing overestimation of $\kappa_{\rm L}$ by HA+3ph for most functionals. On the other hand, LDA provides a clearer demonstration of the alleviation of the $\kappa_{\rm L}$ underestimation from HA+3ph to SCPH+3ph. Notably, the overestimation of $\kappa_{\rm L}$ for BAs and ZnTe is slightly reduced due to the slight decrease in phonon frequency caused by SCPH, as shown in Fig.~\ref{fig:phonon}(b). The effects of temperature on $\kappa_{\rm L}$ can be more accurately assessed for each compound by calculating the ratio between $\kappa_{\rm 3ph}^{\rm SCPH}$ and $\kappa_{\rm 3ph}^{\rm HA}$. As illustrated in Fig.~\ref{fig:ratio}(a), the majority ratios of $\kappa_{\rm SCPH}^{\rm 3ph}/\kappa_{\rm 3ph}^{\rm HA}$ exceed 1, indicating an enlargement of $\kappa_{\rm L}$ from HA+3ph to SCPH+3ph. Notably, compounds with the rocksalt structure exhibit larger ratios compared to those with the zincblende structure, consistent with previous findings using the PBE functional~\cite{RN37}, attributed to the lower coordination number and weaker anharmonicity of the zincblende structure~\cite{doi:10.1021/acs.chemmater.6b04179}. Furthermore, the degree of enhancement varies with the functional used for a specific compound. For instance, the $\kappa_{\rm 3ph}^{\rm SCPH}/\kappa_{\rm 3ph}^{\rm HA}$ ratios for NaCl are 1.67, 1.0, 1.48, 1.28, 1.68, 2.33, 0.79, 1.25 for LDA, PBE, PBEsol, optB86b, revTPSS, SCAN, rSCAN, and r$^2$SCAN, respectively.

Notably, all of these functionals exhibit a large $\kappa_{\rm 3ph}^{\rm SCPH}/\kappa_{\rm 3ph}^{\rm HA}$ ratio for BaO, but a smaller ratio for other alkaline earth metal oxides $M$O ($M$ = Mg, Sr, and Ca). This difference is likely attributed to the greater phonon frequency shift from SCPH to HA in BaO compared to the other $M$O compounds, as shown in Fig.~\ref{fig:phonon}(b). Except SCAN-related functionals, all the other functionals demonstrate a similar trend in the $\kappa_{\rm 3ph}^{\rm SCPH}/\kappa_{\rm 3ph}^{\rm HA}$ ratio for compounds with the zincblende structure, where most compounds have a ratio of around 1. Conversely, SCAN-related functionals display more variability in this ratio. Specifically, SCAN and rSCAN indicate a ratio greater than 1 for ZnSe, while rSCAN and r$^2$SCAN show ratios greater than 1 for BAs and ZnTe, respectively. It is worth noting that rSCAN deviates from the trend observed in other functionals when calculating the $\kappa_{\rm 3ph}^{\rm SCPH}/\kappa_{\rm 3ph}^{\rm HA}$ ratio of NaCl in rocksalt compounds: while all other functionals have a ratio of around 1, rSCAN yields a ratio of 0.79. This suggests that SCAN-related functionals can not provide consistent results when calculating $\kappa_{\rm 3ph}^{\rm HA}$ and $\kappa_{\rm 3ph}^{\rm SCPH}$.

Fig.~\ref{fig:ha+scph}(c) displays the calculated $\kappa_{\rm L}$ at the SCPH+3,4ph level, which represents the most accurate method utilized in this study. The maximum and minimum MRAEs are 35.06 \% with rSCAN and 20.03 \% with PBEsol, respectively. A comparison with Fig.~\ref{fig:ha+scph}(b) reveals a notable decrease in MRAEs attributed to additional phonon-phonon scattering from four-phonon interactions, which is particularly significant for systems with relatively weak three-phonon scattering like BAs~\cite{PhysRevB.96.161201}. It is evident that four-phonon scattering significantly impacts the calculated $\kappa_{\rm L}$ of the compounds with the zincblende structure, contrary to the effect of SCPH as discussed earlier. In stark contrast to the results from HA+3ph and SCPH+3ph, the LDA-calculated $\kappa_{\rm 3,4ph}^{\rm SCPH}$ for all the examined compounds in this study are lower than the experimental $\kappa_{\rm L}$, with larger relative errors for the compounds with the rocksalt structure. This discrepancy is attributed to the overbinding issue of LDA, which results in smaller volumes (shorter bond lengths) compared to these at room temperature, see Fig.~\textcolor{magenta}{S14}, while the experimental lattice constants at room temperature are employed in this work, which leads to weaker chemical bonds in our LDA calculations. A similar trend is observed in PBEsol, which exhibits slightly larger equilibrium volumes than LDA at 0 K but still smaller than room-temperature values for most compounds, see Fig.~\textcolor{magenta}{S14}. The agreement with experimental lattice constants will be significantly improved if the thermal expansion is considered~\cite{PhysRevB.108.024306}. Interestingly, despite PBE-optimized volumes at 0 K being larger than experimental volumes at room temperature (see Fig.~\textcolor{magenta}{S14}), resulting in enhanced chemical bonds when utilizing room-temperature lattice constants, PBE only overestimates the $\kappa_{\rm L}$ of three compounds (BaO, MgO, and PbTe). This implies a more complex relationship between bond length and $\kappa_{\rm L}$ across various XC functionals. Among these functionals, only LDA, PBEsol, and SCAN barely address the issue of overestimation for BaO, showing the best agreement with experimental values among these functional~\cite{simoncelli2019unified}. While PBE, optB86b, and revTPSS largely overestimate the $\kappa_{\rm 3,4ph}^{\rm SCPH}$ of PbTe, other functionals achieve good agreement with the experimental value if no imaginary frequencies are present at the SCPH level. Additionally, we observed that certain functionals yield significantly lower $\kappa_{\rm 3,4ph}^{\rm SCPH}$ values for specific compounds compared to other XC functionals. Examples include SCAN for CaO, LDA for PbS, revTPSS and rSCAN for BAs, SCAN for ZnSe, and r$^2$SCAN for ZnTe. Conversely, all functionals largely underestimate the $\kappa_{\rm L}$ of InAs and InSb except SCAN, suggesting potential inaccuracies in the experimental values or the influence of other factors such as lattice constants. Overall, the performance of the PBEsol functional is notable in most of these compounds, with a slightly smaller relative error observed for compounds with a rocksalt structure compared to those with a zincblende structure.

\begin{figure}[tph!]
	\includegraphics[clip,width=1.0\linewidth]{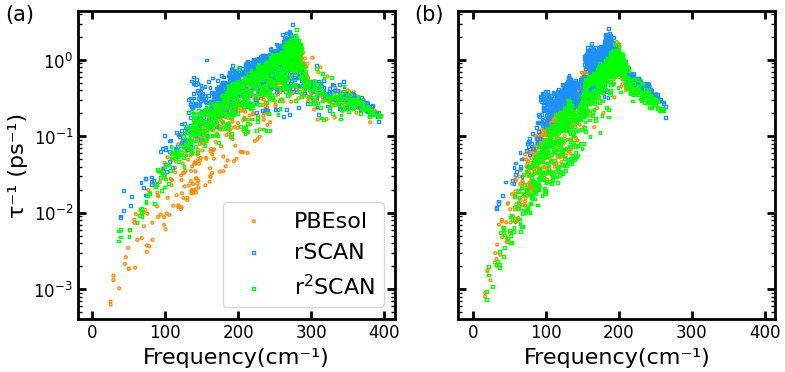}
	\caption{(Color online) (a) and (b) are the four-phonon scattering rates calculated using PBEsol, rSCAN, and r$^2$SCAN for NaF and NaCl, respectively.}
	\label{fig:NaF-NaCl}
\end{figure}

The effect of four-phonon scattering on $\kappa_{\rm L}$ can be assessed by examining the ratio of $\kappa_{\rm 3,4ph}^{\rm SCPH}/\kappa_{\rm 3ph}^{\rm SCPH}$. Analysis presented in Fig.~\ref{fig:ratio}(b) reveals that the compounds with the zincblende structure are more significantly affected by the four-phonon scattering compared to those with the rocksalt one, with the exception of SiC, BP, and GaAs across all functionals. This observation suggests that four-phonon scattering is relatively weaker in rocksalt compounds because it has strong three-phonon scattering. Surprisingly, all the functionals find that the ratios of $\kappa_{\rm 3,4ph}^{\rm SCPH}/\kappa_{\rm 3ph}^{\rm SCPH}$ for InSb, InAs, and ZnTe are comparable or even smaller than that of BAs, a well-known compound with strong four-phonon scattering~\cite{PhysRevB.96.161201}, indicating a more pronounced four-phonon interaction effect. In contrast to other functionals, rSCAN predicts that the rocksalt compounds NaF and NaCl also exhibit strong four-phonon scattering, resulting in a significant underestimation of $\kappa_{\rm L}$. A similar result is predicted by r$^2$SCAN for NaF, but not for NaCl. The four-phonon scattering rates of these two compounds as predicted by rSCAN are illustrated in Fig.~\ref{fig:NaF-NaCl}. When compared to the PBEsol functional, rSCAN demonstrates significantly stronger four-phonon scattering in the low-frequency region for both NaF and NaCl. In contrast, r$^2$SCAN only predicts high four-phonon scattering in NaF, aligning with the observed trend in $\kappa_{\rm L}$.

The $\kappa_{\rm 3,4ph}^{\rm SCPH}/\kappa_{\rm 3ph}^{\rm HA}$ ratio provides valuable insight into the relative contributions of phonon renormalization at room temperature and the four-phonon scattering in the calculation of $\kappa_{\rm L}$. A $\kappa_{\rm 3,4ph}^{\rm SCPH}/\kappa_{\rm 3ph}^{\rm HA}$ ratio of 1 indicates a perfect cancellation between these two effects, while a ratio of less than 1 suggests that the reduction of $\kappa_{\rm L}$ due to four-phonon scattering outweighs the enhancement from phonon renormalization, although the latter can occasionally lead to a decrease in $\kappa_{\rm L}$ as well. Analysis of the data presented in Fig.~\ref{fig:ratio}(c) indicates that all zincblende compounds demonstrate a $\kappa_{\rm 3,4ph}^{\rm SCPH}/\kappa_{\rm 3ph}^{\rm HA}$ ratio below 1 across all functionals, which is attributed to the smaller or negative phonon frequency shifts at room temperature and the heightened four-phonon scattering characteristic of the zincblende structure, aligning with previous findings~\cite{RN37}. Among these compounds, SiC and BP stand out as the only zincblende compounds with a $\kappa_{\rm 3,4ph}^{\rm SCPH}/\kappa_{\rm 3ph}^{\rm HA}$ ratio approaching 1. Conversely, all rocksalt compounds exhibit a $\kappa_{\rm 3,4ph}^{\rm SCPH}/\kappa_{\rm 3ph}^{\rm HA}$ ratio close to or greater than 1. Notable exceptions to this trend are primarily observed in NaF and NaCl across nearly all functionals. At the same time, these two compounds are predicted to have abnormally smaller $\kappa_{\rm 3,4ph}^{\rm SCPH}/\kappa_{\rm 3ph}^{\rm HA}$ ratio than other functionals by rSCAN. It is evident that the rocksalt compounds generally display similar values for $\kappa_{\rm 3,4ph}^{\rm SCPH}/\kappa_{\rm 3ph}^{\rm HA}$ ratio, whereas the zincblende compounds show large frustration across these compounds. BaO is identified as a notable case where the inclusion of four-phonon scattering leads to a significant reduction in $\kappa_{\rm L}$, surpassing the overestimation produced by SCPH+3ph.

\subsection{Dependence of lattice thermal conductivity on volume}
\begin{figure}[tph!]
	\includegraphics[clip,width=1.0\linewidth]{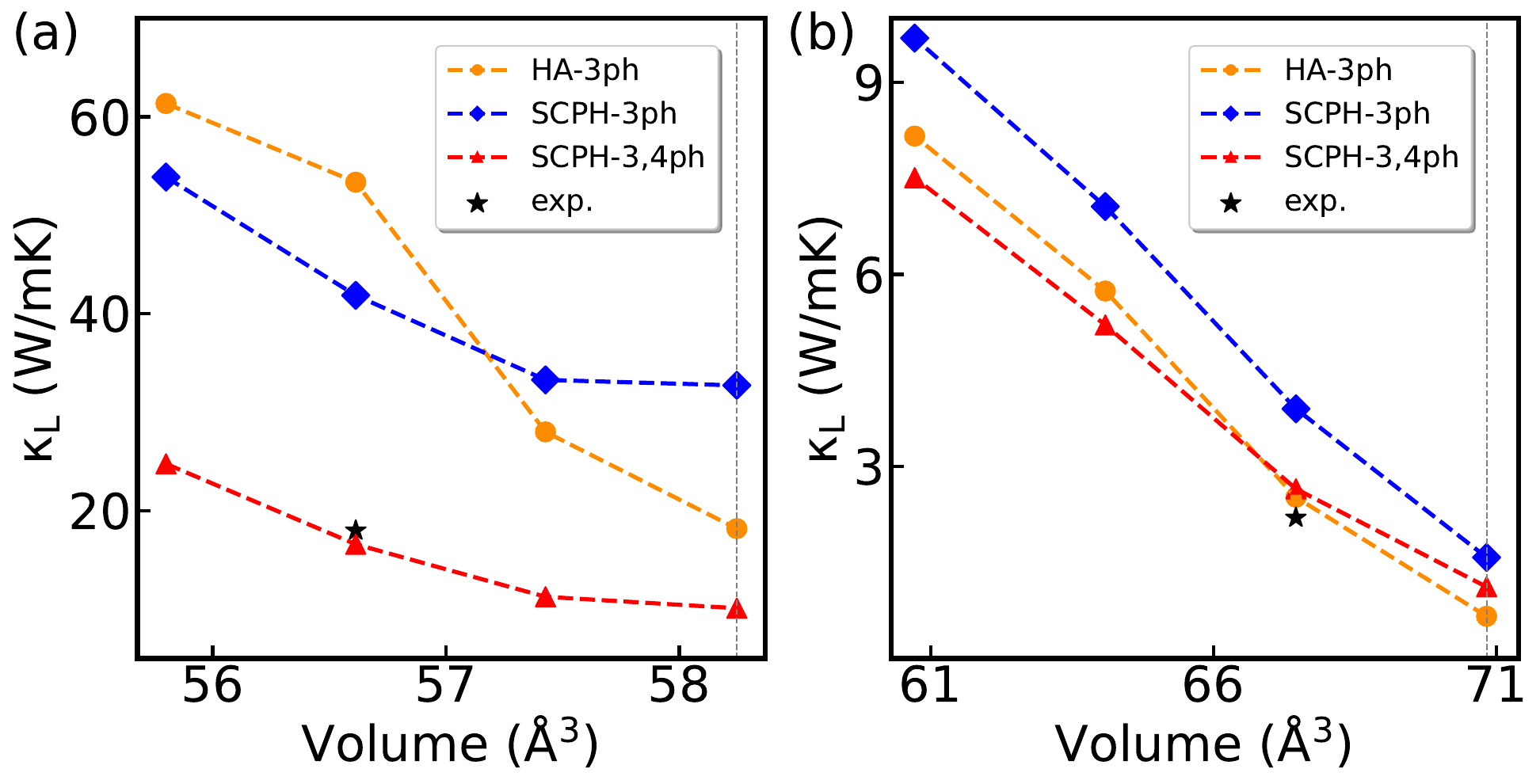}
	\caption{(Color online). (a) and (b) are the PBEsol calculated dependence of $\kappa_{\rm L}$ on the volume for ZnTe and PbTe calculated at three perturbation orders (HA+3ph, SCPH+3ph, and SCPH+3,4ph), respectively. The vertical dash line indicates the PBEsol optimized volume.}
	\label{fig:znte_volume}
\end{figure}

Our previous study demonstrates that there is a significant disparity between the computed room-temperature lattice constants using QHA and the corresponding experimental values for many of these XC functionals~\cite{PhysRevB.108.024306}. However, due to the considerable cost associated with computing thermal expansion through anharmonic approximation for such an extensive dataset, this work employs the room-temperature experimental lattice constants to mitigate the influence of lattice thermal expansion on $\kappa_{\rm L}$. As illustrated in Fig.~\textcolor{magenta}{S14}, most compounds exhibit significant discrepancies between the optimized volumes at 0 K and the experimental ones at room temperature. Consequently, the chemical bonding of these compounds may be either strengthened or weakened by the \textit{smaller} or \textit{larger} volumes in our calculations. A stronger (weaker) chemical bond could result in a higher (lower) $v_{\rm g}$ and, consequently, an increased (decreased) $\kappa_{\rm L}$~\cite{https://doi.org/10.1002/adfm.202108532}. To further elucidate the impact of volume on the $\kappa_{\rm L}$ of these compounds, we present the correlation between $\kappa_{\rm L}$ and volume for two represented compounds ZnTe and PbTe in Fig.~\ref{fig:znte_volume}. As demonstrated above, PBEsol exhibits exceptional performance in computing $\kappa_{\rm L}$ across all perturbation orders. Consequently, PBEsol is utilized to explore the dependence of $\kappa_{\rm L}$ on volume. ZnTe is a zincblende compound, and our calculated $\kappa_{\rm L}$ is notably higher than that reported in a previous study~\cite{RN37}. On the other hand, PbTe serves as a representative of the rocksalt compound with significant anharmonicity. It is observed that the $\kappa_{\rm L}$ increases monotonically as the volume decreases (corresponding to an increase in pressure) for both ZnTe and PbTe. Interestingly, while the $\kappa_{\rm 3,4ph}^{\rm SCPH}$ of ZnTe is nearly linearly dependent on volume, the changes of $\kappa_{\rm 3ph}^{\rm HA}$ and $\kappa_{\rm 3ph}^{\rm SCPH}$ with volume is nonlinear. Additionally, both $\kappa_{\rm 3ph}^{\rm HA}$ and $\kappa_{\rm 3ph}^{\rm SCPH}$ are much higher than $\kappa_{\rm 3,4ph}^{\rm SCPH}$, and the difference between them widens as the volume decreases, suggesting a significant enhancement in four-phonon scattering. Notably, a crossover is observed between $\kappa_{\rm 3ph}^{\rm HA}$ and $\kappa_{\rm 3ph}^{\rm SCPH}$ at a volume close to the experimental one. When the volume is smaller than this critical volume, $\kappa_{\rm 3ph}^{\rm SCPH}$ is lower than $\kappa_{\rm 3ph}^{\rm HA}$. In contrast to ZnTe, the behavior of PbTe is distinct, with $\kappa_{\rm 3ph}^{\rm HA}$, $\kappa_{\rm 3ph}^{\rm SCPH}$, and $\kappa_{\rm 3,4ph}^{\rm SCPH}$ exhibiting a linear increase as the volume decreases, albeit with slightly different slopes. Notably, for any given volume, $\kappa_{\rm 3ph}^{\rm SCPH}$ consistently exceeds $\kappa_{\rm 3ph}^{\rm HA}$. These differences stem from the disparity in the impact of the SCPH effect on rocksalt and zincblende structures.

\section{Conclusions}
In this study, we systematically investigate the performance of eight commonly used and newly proposed XC functionals in the solid community on lattice thermal conductivity based on the phonon gas model with three different perturbation orders. Within the approximation of harmonic phonons and three-phonon scattering, all the XC functionals tend to overestimate $\kappa_{\rm L}$ for almost all the compounds. Among these functionals, revTPSS exhibits the smallest MRAE (29.75 \%), followed by SCAN (30.11 \%), LDA (34.00 \%), and r$^2$SCAN (35.23 \%), whereas PBE (45.51 \%) and rSCAN (49.18 \%) possess the largest MREA among these XC functionals. If the phonon spectrum is renormalized at room temperature using the SCPH method and only include three-phonon scattering (SCPH+3ph), the $\kappa_{\rm L}$ values are further enlarged for nearly all the functionals, due to the phonon frequency shift towards higher energy, resulting in minimal MRAE of 30.57 \% by SCAN and maximal MRAE of 73.11 \% by r$^2$SCAN. However, if the four-phonon scattering is included in the SCPH+3ph as well (SCPH+3,4ph), which is the most accurate method employed in this work, the MRAEs of all the functionals are significantly reduced, achieving the minimal MRAE of 20.03 \% by PBEsol and maximal MRAE of 35.06 \% by rSCAN. Both rSCAN and r$^2$SCAN have improved numerical stability, but only r$^2$SCAN has comparable MARE with PBEsol, implying the importance of adhering to the constraints imposed by SCAN. However, although metaGGA functionals have a higher hierarchy and are computationally more expensive than LDA and GGAs, their accuracy in capturing anharmonicity is slightly inferior to the lower hierarchy functionals, particularly PBEsol. Due to the small difference in phonon spectra and the smaller MRAE in sound velocity compared to the Gr\"uneisen parameter, the primary source of error across these functionals predominantly arises from the difference in dealing with anharmonicity. Our findings not only offer a comprehensive assessment of XC functionals for computing $\kappa_{\rm L}$, but also provide a useful guide for selecting appropriate XC functionals when dealing with anharmonicity. Furthermore, our results highlight the challenges of accurately describing anharmonicity using XC functionals.

\section{ACKNOWLEDGMENTS}
J.W. Z.X. and J.H. acknowledge the support of the National Science Fundation of China (12374024). Y. X. acknowledges support from the US National Science Foundation through award 2317008. The computing resource was supported by USTB MatCom of Beijing Advanced Innovation Center for Materials Genome Engineering.

\bibliography{ref}

\begin{thebibliography}{89}%
\makeatletter
\providecommand \@ifxundefined [1]{%
 \@ifx{#1\undefined}
}%
\providecommand \@ifnum [1]{%
 \ifnum #1\expandafter \@firstoftwo
 \else \expandafter \@secondoftwo
 \fi
}%
\providecommand \@ifx [1]{%
 \ifx #1\expandafter \@firstoftwo
 \else \expandafter \@secondoftwo
 \fi
}%
\providecommand \natexlab [1]{#1}%
\providecommand \enquote  [1]{``#1''}%
\providecommand \bibnamefont  [1]{#1}%
\providecommand \bibfnamefont [1]{#1}%
\providecommand \citenamefont [1]{#1}%
\providecommand \href@noop [0]{\@secondoftwo}%
\providecommand \href [0]{\begingroup \@sanitize@url \@href}%
\providecommand \@href[1]{\@@startlink{#1}\@@href}%
\providecommand \@@href[1]{\endgroup#1\@@endlink}%
\providecommand \@sanitize@url [0]{\catcode `\\12\catcode `\$12\catcode
  `\&12\catcode `\#12\catcode `\^12\catcode `\_12\catcode `\%12\relax}%
\providecommand \@@startlink[1]{}%
\providecommand \@@endlink[0]{}%
\providecommand \url  [0]{\begingroup\@sanitize@url \@url }%
\providecommand \@url [1]{\endgroup\@href {#1}{\urlprefix }}%
\providecommand \urlprefix  [0]{URL }%
\providecommand \Eprint [0]{\href }%
\providecommand \doibase [0]{https://doi.org/}%
\providecommand \selectlanguage [0]{\@gobble}%
\providecommand \bibinfo  [0]{\@secondoftwo}%
\providecommand \bibfield  [0]{\@secondoftwo}%
\providecommand \translation [1]{[#1]}%
\providecommand \BibitemOpen [0]{}%
\providecommand \bibitemStop [0]{}%
\providecommand \bibitemNoStop [0]{.\EOS\space}%
\providecommand \EOS [0]{\spacefactor3000\relax}%
\providecommand \BibitemShut  [1]{\csname bibitem#1\endcsname}%
\let\auto@bib@innerbib\@empty
\bibitem [{\citenamefont {Moore}\ and\ \citenamefont
  {Shi}(2014)}]{MOORE2014163}%
  \BibitemOpen
  \bibfield  {author} {\bibinfo {author} {\bibfnamefont {A.~L.}\ \bibnamefont
  {Moore}}\ and\ \bibinfo {author} {\bibfnamefont {L.}~\bibnamefont {Shi}},\
  }\bibfield  {title} {\bibinfo {title} {Emerging challenges and materials for
  thermal management of electronics},\ }\href
  {https://doi.org/https://doi.org/10.1016/j.mattod.2014.04.003} {\bibfield
  {journal} {\bibinfo  {journal} {Mater. Today}\ }\textbf {\bibinfo {volume}
  {17}},\ \bibinfo {pages} {163} (\bibinfo {year} {2014})}\BibitemShut
  {NoStop}%
\bibitem [{\citenamefont {Zhang}\ \emph {et~al.}(2021)\citenamefont {Zhang},
  \citenamefont {Wang},\ and\ \citenamefont {Yan}}]{2}%
  \BibitemOpen
  \bibfield  {author} {\bibinfo {author} {\bibfnamefont {Z.}~\bibnamefont
  {Zhang}}, \bibinfo {author} {\bibfnamefont {X.}~\bibnamefont {Wang}},\ and\
  \bibinfo {author} {\bibfnamefont {Y.}~\bibnamefont {Yan}},\ }\bibfield
  {title} {\bibinfo {title} {A review of the state-of-the-art in electronic
  cooling},\ }\href
  {https://doi.org/https://doi.org/10.1016/j.prime.2021.100009} {\bibfield
  {journal} {\bibinfo  {journal} {e-Prime}\ }\textbf {\bibinfo {volume} {1}},\
  \bibinfo {pages} {100009} (\bibinfo {year} {2021})}\BibitemShut {NoStop}%
\bibitem [{\citenamefont {Rowe}(1995)}]{CRC-Handbook}%
  \BibitemOpen
  \bibfield  {author} {\bibinfo {author} {\bibfnamefont {D.~E.}\ \bibnamefont
  {Rowe}},\ }\href {https://doi.org/10.1201/9781420049718} {\emph {\bibinfo
  {title} {CRC Handbook of Thermoelectrics (1st ed.)}}}\ (\bibinfo  {publisher}
  {CRC Press},\ \bibinfo {year} {1995})\BibitemShut {NoStop}%
\bibitem [{\citenamefont {Bell}(2008)}]{bell2008cooling}%
  \BibitemOpen
  \bibfield  {author} {\bibinfo {author} {\bibfnamefont {L.~E.}\ \bibnamefont
  {Bell}},\ }\bibfield  {title} {\bibinfo {title} {Cooling, heating, generating
  power, and recovering waste heat with thermoelectric systems},\ }\href
  {https://doi.org/10.1126/science.1158899} {\bibfield  {journal} {\bibinfo
  {journal} {Science}\ }\textbf {\bibinfo {volume} {321}},\ \bibinfo {pages}
  {1457} (\bibinfo {year} {2008})}\BibitemShut {NoStop}%
\bibitem [{\citenamefont {Mondal}\ \emph {et~al.}(2021)\citenamefont {Mondal},
  \citenamefont {Nuñez}, \citenamefont {Downey},\ and\ \citenamefont {van
  Rooyen}}]{doi:10.1021/acs.iecr.1c00788}%
  \BibitemOpen
  \bibfield  {author} {\bibinfo {author} {\bibfnamefont {K.}~\bibnamefont
  {Mondal}}, \bibinfo {author} {\bibfnamefont {L.~I.}\ \bibnamefont {Nuñez}},
  \bibinfo {author} {\bibfnamefont {C.~M.}\ \bibnamefont {Downey}},\ and\
  \bibinfo {author} {\bibfnamefont {I.~J.}\ \bibnamefont {van Rooyen}},\
  }\bibfield  {title} {\bibinfo {title} {Thermal barrier coatings overview:
  Design, manufacturing, and applications in high-temperature industries},\
  }\href {https://doi.org/10.1021/acs.iecr.1c00788} {\bibfield  {journal}
  {\bibinfo  {journal} {Ind. Eng. Chem. Res.}\ }\textbf {\bibinfo {volume}
  {60}},\ \bibinfo {pages} {6061} (\bibinfo {year} {2021})}\BibitemShut
  {NoStop}%
\bibitem [{\citenamefont {Padture}\ \emph {et~al.}(2002)\citenamefont
  {Padture}, \citenamefont {Gell},\ and\ \citenamefont
  {Jordan}}]{padture2002thermal}%
  \BibitemOpen
  \bibfield  {author} {\bibinfo {author} {\bibfnamefont {N.~P.}\ \bibnamefont
  {Padture}}, \bibinfo {author} {\bibfnamefont {M.}~\bibnamefont {Gell}},\ and\
  \bibinfo {author} {\bibfnamefont {E.~H.}\ \bibnamefont {Jordan}},\ }\bibfield
   {title} {\bibinfo {title} {Thermal barrier coatings for gas-turbine engine
  applications},\ }\href {https://doi.org/10.1126/science.1068609} {\bibfield
  {journal} {\bibinfo  {journal} {Science}\ }\textbf {\bibinfo {volume}
  {296}},\ \bibinfo {pages} {280} (\bibinfo {year} {2002})}\BibitemShut
  {NoStop}%
\bibitem [{\citenamefont {Clarke}\ and\ \citenamefont {Phillpot}(2005)}]{1}%
  \BibitemOpen
  \bibfield  {author} {\bibinfo {author} {\bibfnamefont {D.~R.}\ \bibnamefont
  {Clarke}}\ and\ \bibinfo {author} {\bibfnamefont {S.~R.}\ \bibnamefont
  {Phillpot}},\ }\bibfield  {title} {\bibinfo {title} {Thermal barrier coating
  materials},\ }\href
  {https://doi.org/https://doi.org/10.1016/S1369-7021(05)70934-2} {\bibfield
  {journal} {\bibinfo  {journal} {Mater. Today}\ }\textbf {\bibinfo {volume}
  {8}},\ \bibinfo {pages} {22} (\bibinfo {year} {2005})}\BibitemShut {NoStop}%
\bibitem [{\citenamefont {Lindsay}\ \emph {et~al.}(2013)\citenamefont
  {Lindsay}, \citenamefont {Broido},\ and\ \citenamefont
  {Reinecke}}]{PhysRevLett.111.025901}%
  \BibitemOpen
  \bibfield  {author} {\bibinfo {author} {\bibfnamefont {L.}~\bibnamefont
  {Lindsay}}, \bibinfo {author} {\bibfnamefont {D.~A.}\ \bibnamefont
  {Broido}},\ and\ \bibinfo {author} {\bibfnamefont {T.~L.}\ \bibnamefont
  {Reinecke}},\ }\bibfield  {title} {\bibinfo {title} {First-principles
  determination of ultrahigh thermal conductivity of boron arsenide: A
  competitor for diamond?},\ }\href
  {https://doi.org/10.1103/PhysRevLett.111.025901} {\bibfield  {journal}
  {\bibinfo  {journal} {Phys. Rev. Lett.}\ }\textbf {\bibinfo {volume} {111}},\
  \bibinfo {pages} {025901} (\bibinfo {year} {2013})}\BibitemShut {NoStop}%
\bibitem [{\citenamefont {Yang}\ \emph {et~al.}(2019)\citenamefont {Yang},
  \citenamefont {Feng}, \citenamefont {Li},\ and\ \citenamefont
  {Ruan}}]{PhysRevB.100.245203}%
  \BibitemOpen
  \bibfield  {author} {\bibinfo {author} {\bibfnamefont {X.}~\bibnamefont
  {Yang}}, \bibinfo {author} {\bibfnamefont {T.}~\bibnamefont {Feng}}, \bibinfo
  {author} {\bibfnamefont {J.}~\bibnamefont {Li}},\ and\ \bibinfo {author}
  {\bibfnamefont {X.}~\bibnamefont {Ruan}},\ }\bibfield  {title} {\bibinfo
  {title} {Stronger role of four-phonon scattering than three-phonon scattering
  in thermal conductivity of iii-v semiconductors at room temperature},\ }\href
  {https://doi.org/10.1103/PhysRevB.100.245203} {\bibfield  {journal} {\bibinfo
   {journal} {Phys. Rev. B}\ }\textbf {\bibinfo {volume} {100}},\ \bibinfo
  {pages} {245203} (\bibinfo {year} {2019})}\BibitemShut {NoStop}%
\bibitem [{\citenamefont {Xia}\ \emph {et~al.}(2020{\natexlab{a}})\citenamefont
  {Xia}, \citenamefont {Hegde}, \citenamefont {Pal}, \citenamefont {Hua},
  \citenamefont {Gaines}, \citenamefont {Patel}, \citenamefont {He},
  \citenamefont {Aykol},\ and\ \citenamefont {Wolverton}}]{RN37}%
  \BibitemOpen
  \bibfield  {author} {\bibinfo {author} {\bibfnamefont {Y.}~\bibnamefont
  {Xia}}, \bibinfo {author} {\bibfnamefont {V.~I.}\ \bibnamefont {Hegde}},
  \bibinfo {author} {\bibfnamefont {K.}~\bibnamefont {Pal}}, \bibinfo {author}
  {\bibfnamefont {X.}~\bibnamefont {Hua}}, \bibinfo {author} {\bibfnamefont
  {D.}~\bibnamefont {Gaines}}, \bibinfo {author} {\bibfnamefont
  {S.}~\bibnamefont {Patel}}, \bibinfo {author} {\bibfnamefont
  {J.}~\bibnamefont {He}}, \bibinfo {author} {\bibfnamefont {M.}~\bibnamefont
  {Aykol}},\ and\ \bibinfo {author} {\bibfnamefont {C.}~\bibnamefont
  {Wolverton}},\ }\bibfield  {title} {\bibinfo {title} {High-throughput study
  of lattice thermal conductivity in binary rocksalt and zinc blende compounds
  including higher-order anharmonicity,}\ }\href {https://doi.org/Phys. Rev. X
  10, 041029} {Phys. Rev. X 10, 041029} (\bibinfo {year}
  {2020}{\natexlab{a}})\BibitemShut {NoStop}%
\bibitem [{\citenamefont {Green}(1954)}]{10.1063/1.1740082}%
  \BibitemOpen
  \bibfield  {author} {\bibinfo {author} {\bibfnamefont {M.~S.}\ \bibnamefont
  {Green}},\ }\bibfield  {title} {\bibinfo {title} {{Markoff Random Processes
  and the Statistical Mechanics of Time‐Dependent Phenomena. II. Irreversible
  Processes in Fluids}},\ }\href {https://doi.org/10.1063/1.1740082} {\bibfield
   {journal} {\bibinfo  {journal} {J. Chem. Phys.}\ }\textbf {\bibinfo {volume}
  {22}},\ \bibinfo {pages} {398} (\bibinfo {year} {1954})}\BibitemShut
  {NoStop}%
\bibitem [{\citenamefont {Evans}\ and\ \citenamefont
  {Morriss}(2008)}]{Evans_Morriss_2008}%
  \BibitemOpen
  \bibfield  {author} {\bibinfo {author} {\bibfnamefont {D.~J.}\ \bibnamefont
  {Evans}}\ and\ \bibinfo {author} {\bibfnamefont {G.}~\bibnamefont
  {Morriss}},\ }\href
  {https://www.cambridge.org/core/books/statistical-mechanics-of-nonequilibrium-liquids/E97F04B0D7D423540AA3F9B91F155E7D}
  {\emph {\bibinfo {title} {Statistical Mechanics of Nonequilibrium
  Liquids}}},\ \bibinfo {edition} {2nd}\ ed.\ (\bibinfo  {publisher} {Cambridge
  University Press},\ \bibinfo {year} {2008})\BibitemShut {NoStop}%
\bibitem [{\citenamefont {Klemens}(1958)}]{KLEMENS19581}%
  \BibitemOpen
  \bibfield  {author} {\bibinfo {author} {\bibfnamefont {P.~G.}\ \bibnamefont
  {Klemens}},\ }\bibfield  {title} {\bibinfo {title} {Thermal conductivity and
  lattice vibrational modes},\ }\href
  {https://api.semanticscholar.org/CorpusID:118768509} {\bibfield  {journal}
  {\bibinfo  {journal} {J. Phys. C: Solid State Phys.}\ }\textbf {\bibinfo
  {volume} {7}},\ \bibinfo {pages} {1} (\bibinfo {year} {1958})}\BibitemShut
  {NoStop}%
\bibitem [{\citenamefont {Puligheddu}\ \emph {et~al.}(2019)\citenamefont
  {Puligheddu}, \citenamefont {Xia}, \citenamefont {Chan},\ and\ \citenamefont
  {Galli}}]{PhysRevMaterials.3.085401}%
  \BibitemOpen
  \bibfield  {author} {\bibinfo {author} {\bibfnamefont {M.}~\bibnamefont
  {Puligheddu}}, \bibinfo {author} {\bibfnamefont {Y.}~\bibnamefont {Xia}},
  \bibinfo {author} {\bibfnamefont {M.}~\bibnamefont {Chan}},\ and\ \bibinfo
  {author} {\bibfnamefont {G.}~\bibnamefont {Galli}},\ }\bibfield  {title}
  {\bibinfo {title} {Computational prediction of lattice thermal conductivity:
  A comparison of molecular dynamics and boltzmann transport approaches},\
  }\href {https://doi.org/10.1103/PhysRevMaterials.3.085401} {\bibfield
  {journal} {\bibinfo  {journal} {Phys. Rev. Mater.}\ }\textbf {\bibinfo
  {volume} {3}},\ \bibinfo {pages} {085401} (\bibinfo {year}
  {2019})}\BibitemShut {NoStop}%
\bibitem [{\citenamefont {Feng}\ \emph {et~al.}(2017)\citenamefont {Feng},
  \citenamefont {Lindsay},\ and\ \citenamefont {Ruan}}]{PhysRevB.96.161201}%
  \BibitemOpen
  \bibfield  {author} {\bibinfo {author} {\bibfnamefont {T.}~\bibnamefont
  {Feng}}, \bibinfo {author} {\bibfnamefont {L.}~\bibnamefont {Lindsay}},\ and\
  \bibinfo {author} {\bibfnamefont {X.}~\bibnamefont {Ruan}},\ }\bibfield
  {title} {\bibinfo {title} {Four-phonon scattering significantly reduces
  intrinsic thermal conductivity of solids},\ }\href
  {https://doi.org/10.1103/PhysRevB.96.161201} {\bibfield  {journal} {\bibinfo
  {journal} {Phys. Rev. B}\ }\textbf {\bibinfo {volume} {96}},\ \bibinfo
  {pages} {161201} (\bibinfo {year} {2017})}\BibitemShut {NoStop}%
\bibitem [{\citenamefont {Han}\ \emph {et~al.}(2022)\citenamefont {Han},
  \citenamefont {Yang}, \citenamefont {Li}, \citenamefont {Feng},\ and\
  \citenamefont {Ruan}}]{HAN2022108179}%
  \BibitemOpen
  \bibfield  {author} {\bibinfo {author} {\bibfnamefont {Z.}~\bibnamefont
  {Han}}, \bibinfo {author} {\bibfnamefont {X.}~\bibnamefont {Yang}}, \bibinfo
  {author} {\bibfnamefont {W.}~\bibnamefont {Li}}, \bibinfo {author}
  {\bibfnamefont {T.}~\bibnamefont {Feng}},\ and\ \bibinfo {author}
  {\bibfnamefont {X.}~\bibnamefont {Ruan}},\ }\bibfield  {title} {\bibinfo
  {title} {Fourphonon: An extension module to shengbte for computing
  four-phonon scattering rates and thermal conductivity},\ }\href
  {https://doi.org/https://doi.org/10.1016/j.cpc.2021.108179} {\bibfield
  {journal} {\bibinfo  {journal} {Comput. Phys. Commun.}\ }\textbf {\bibinfo
  {volume} {270}},\ \bibinfo {pages} {108179} (\bibinfo {year}
  {2022})}\BibitemShut {NoStop}%
\bibitem [{\citenamefont {Xia}(2018)}]{10.1063/1.5040887}%
  \BibitemOpen
  \bibfield  {author} {\bibinfo {author} {\bibfnamefont {Y.}~\bibnamefont
  {Xia}},\ }\bibfield  {title} {\bibinfo {title} {{Revisiting lattice thermal
  transport in PbTe: The crucial role of quartic anharmonicity}},\ }\href
  {https://doi.org/10.1063/1.5040887} {\bibfield  {journal} {\bibinfo
  {journal} {Appl. Phys. Lett.}\ }\textbf {\bibinfo {volume} {113}},\ \bibinfo
  {pages} {073901} (\bibinfo {year} {2018})}\BibitemShut {NoStop}%
\bibitem [{\citenamefont {Perdew}\ and\ \citenamefont
  {Schmidt}(2001)}]{10.1063/1.1390175}%
  \BibitemOpen
  \bibfield  {author} {\bibinfo {author} {\bibfnamefont {J.~P.}\ \bibnamefont
  {Perdew}}\ and\ \bibinfo {author} {\bibfnamefont {K.}~\bibnamefont
  {Schmidt}},\ }\bibfield  {title} {\bibinfo {title} {{Jacob’s ladder of
  density functional approximations for the exchange-correlation energy}},\
  }\href {https://doi.org/10.1063/1.1390175} {\bibfield  {journal} {\bibinfo
  {journal} {AIP Conf. Proc.}\ }\textbf {\bibinfo {volume} {577}},\ \bibinfo
  {pages} {1} (\bibinfo {year} {2001})}\BibitemShut {NoStop}%
\bibitem [{\citenamefont {Zhou}\ \emph {et~al.}(2023)\citenamefont {Zhou},
  \citenamefont {Zhou}, \citenamefont {Hua}, \citenamefont {Bawane},\ and\
  \citenamefont {Feng}}]{10.1063/5.0173762}%
  \BibitemOpen
  \bibfield  {author} {\bibinfo {author} {\bibfnamefont {H.}~\bibnamefont
  {Zhou}}, \bibinfo {author} {\bibfnamefont {S.}~\bibnamefont {Zhou}}, \bibinfo
  {author} {\bibfnamefont {Z.}~\bibnamefont {Hua}}, \bibinfo {author}
  {\bibfnamefont {K.}~\bibnamefont {Bawane}},\ and\ \bibinfo {author}
  {\bibfnamefont {T.}~\bibnamefont {Feng}},\ }\bibfield  {title} {\bibinfo
  {title} {{Extreme sensitivity of higher-order interatomic force constants and
  thermal conductivity to the energy surface roughness of exchange-correlation
  functionals}},\ }\href {https://doi.org/10.1063/5.0173762} {\bibfield
  {journal} {\bibinfo  {journal} {Appl. Phys. Lett.}\ }\textbf {\bibinfo
  {volume} {123}},\ \bibinfo {pages} {192201} (\bibinfo {year}
  {2023})}\BibitemShut {NoStop}%
\bibitem [{\citenamefont {Mardirossian}\ and\ \citenamefont
  {Head-Gordon}(2017)}]{200}%
  \BibitemOpen
  \bibfield  {author} {\bibinfo {author} {\bibfnamefont {N.}~\bibnamefont
  {Mardirossian}}\ and\ \bibinfo {author} {\bibfnamefont {M.}~\bibnamefont
  {Head-Gordon}},\ }\bibfield  {title} {\bibinfo {title} {Thirty years of
  density functional theory in computational chemistry: an overview and
  extensive assessment of 200 density functionals},\ }\href
  {https://doi.org/10.1080/00268976.2017.1333644} {\bibfield  {journal}
  {\bibinfo  {journal} {Mol. Phys.}\ }\textbf {\bibinfo {volume} {115}},\
  \bibinfo {pages} {2315} (\bibinfo {year} {2017})}\BibitemShut {NoStop}%
\bibitem [{\citenamefont {Kohn}\ and\ \citenamefont
  {Sham}(1965)}]{kohn1965self}%
  \BibitemOpen
  \bibfield  {author} {\bibinfo {author} {\bibfnamefont {W.}~\bibnamefont
  {Kohn}}\ and\ \bibinfo {author} {\bibfnamefont {L.~J.}\ \bibnamefont
  {Sham}},\ }\bibfield  {title} {\bibinfo {title} {Self-consistent equations
  including exchange and correlation effects},\ }\href
  {https://doi.org/doi.org/10.1103/PhysRev.140.A1133} {\bibfield  {journal}
  {\bibinfo  {journal} {Phys. Rev.}\ }\textbf {\bibinfo {volume} {140}},\
  \bibinfo {pages} {A1133} (\bibinfo {year} {1965})}\BibitemShut {NoStop}%
\bibitem [{\citenamefont {Perdew}\ \emph {et~al.}(1996)\citenamefont {Perdew},
  \citenamefont {Burke},\ and\ \citenamefont
  {Ernzerhof}}]{PhysRevLett.77.3865}%
  \BibitemOpen
  \bibfield  {author} {\bibinfo {author} {\bibfnamefont {J.~P.}\ \bibnamefont
  {Perdew}}, \bibinfo {author} {\bibfnamefont {K.}~\bibnamefont {Burke}},\ and\
  \bibinfo {author} {\bibfnamefont {M.}~\bibnamefont {Ernzerhof}},\ }\bibfield
  {title} {\bibinfo {title} {Generalized gradient approximation made simple},\
  }\href {https://doi.org/10.1103/PhysRevLett.77.3865} {\bibfield  {journal}
  {\bibinfo  {journal} {Phys. Rev. Lett.}\ }\textbf {\bibinfo {volume} {77}},\
  \bibinfo {pages} {3865} (\bibinfo {year} {1996})}\BibitemShut {NoStop}%
\bibitem [{\citenamefont {Perdew}\ \emph {et~al.}(2008)\citenamefont {Perdew},
  \citenamefont {Ruzsinszky}, \citenamefont {Csonka}, \citenamefont {Vydrov},
  \citenamefont {Scuseria}, \citenamefont {Constantin}, \citenamefont {Zhou},\
  and\ \citenamefont {Burke}}]{PBEsol}%
  \BibitemOpen
  \bibfield  {author} {\bibinfo {author} {\bibfnamefont {J.~P.}\ \bibnamefont
  {Perdew}}, \bibinfo {author} {\bibfnamefont {A.}~\bibnamefont {Ruzsinszky}},
  \bibinfo {author} {\bibfnamefont {G.~I.}\ \bibnamefont {Csonka}}, \bibinfo
  {author} {\bibfnamefont {O.~A.}\ \bibnamefont {Vydrov}}, \bibinfo {author}
  {\bibfnamefont {G.~E.}\ \bibnamefont {Scuseria}}, \bibinfo {author}
  {\bibfnamefont {L.~A.}\ \bibnamefont {Constantin}}, \bibinfo {author}
  {\bibfnamefont {X.}~\bibnamefont {Zhou}},\ and\ \bibinfo {author}
  {\bibfnamefont {K.}~\bibnamefont {Burke}},\ }\bibfield  {title} {\bibinfo
  {title} {Restoring the density-gradient expansion for exchange in solids and
  surfaces},\ }\href {https://doi.org/10.1103/PhysRevLett.100.136406}
  {\bibfield  {journal} {\bibinfo  {journal} {Phys. Rev. Lett.}\ }\textbf
  {\bibinfo {volume} {100}},\ \bibinfo {pages} {136406} (\bibinfo {year}
  {2008})}\BibitemShut {NoStop}%
\bibitem [{\citenamefont {Perdew}\ \emph {et~al.}(2009)\citenamefont {Perdew},
  \citenamefont {Ruzsinszky}, \citenamefont {Csonka}, \citenamefont
  {Constantin},\ and\ \citenamefont {Sun}}]{PhysRevLett.103.026403}%
  \BibitemOpen
  \bibfield  {author} {\bibinfo {author} {\bibfnamefont {J.~P.}\ \bibnamefont
  {Perdew}}, \bibinfo {author} {\bibfnamefont {A.}~\bibnamefont {Ruzsinszky}},
  \bibinfo {author} {\bibfnamefont {G.~I.}\ \bibnamefont {Csonka}}, \bibinfo
  {author} {\bibfnamefont {L.~A.}\ \bibnamefont {Constantin}},\ and\ \bibinfo
  {author} {\bibfnamefont {J.}~\bibnamefont {Sun}},\ }\bibfield  {title}
  {\bibinfo {title} {Workhorse semilocal density functional for condensed
  matter physics and quantum chemistry},\ }\href
  {https://doi.org/10.1103/PhysRevLett.103.026403} {\bibfield  {journal}
  {\bibinfo  {journal} {Phys. Rev. Lett.}\ }\textbf {\bibinfo {volume} {103}},\
  \bibinfo {pages} {026403} (\bibinfo {year} {2009})}\BibitemShut {NoStop}%
\bibitem [{\citenamefont {Sun}\ \emph {et~al.}(2015)\citenamefont {Sun},
  \citenamefont {Ruzsinszky},\ and\ \citenamefont
  {Perdew}}]{PhysRevLett.115.036402}%
  \BibitemOpen
  \bibfield  {author} {\bibinfo {author} {\bibfnamefont {J.}~\bibnamefont
  {Sun}}, \bibinfo {author} {\bibfnamefont {A.}~\bibnamefont {Ruzsinszky}},\
  and\ \bibinfo {author} {\bibfnamefont {J.~P.}\ \bibnamefont {Perdew}},\
  }\bibfield  {title} {\bibinfo {title} {Strongly constrained and appropriately
  normed semilocal density functional},\ }\href
  {https://doi.org/10.1103/PhysRevLett.115.036402} {\bibfield  {journal}
  {\bibinfo  {journal} {Phys. Rev. Lett.}\ }\textbf {\bibinfo {volume} {115}},\
  \bibinfo {pages} {036402} (\bibinfo {year} {2015})}\BibitemShut {NoStop}%
\bibitem [{\citenamefont {Sun}\ \emph {et~al.}(2016{\natexlab{a}})\citenamefont
  {Sun}, \citenamefont {Remsing}, \citenamefont {Zhang}, \citenamefont {Sun},
  \citenamefont {Ruzsinszky}, \citenamefont {Peng}, \citenamefont {Yang},
  \citenamefont {Paul}, \citenamefont {Waghmare}, \citenamefont {Wu} \emph
  {et~al.}}]{sun2016accurate}%
  \BibitemOpen
  \bibfield  {author} {\bibinfo {author} {\bibfnamefont {J.}~\bibnamefont
  {Sun}}, \bibinfo {author} {\bibfnamefont {R.~C.}\ \bibnamefont {Remsing}},
  \bibinfo {author} {\bibfnamefont {Y.}~\bibnamefont {Zhang}}, \bibinfo
  {author} {\bibfnamefont {Z.}~\bibnamefont {Sun}}, \bibinfo {author}
  {\bibfnamefont {A.}~\bibnamefont {Ruzsinszky}}, \bibinfo {author}
  {\bibfnamefont {H.}~\bibnamefont {Peng}}, \bibinfo {author} {\bibfnamefont
  {Z.}~\bibnamefont {Yang}}, \bibinfo {author} {\bibfnamefont {A.}~\bibnamefont
  {Paul}}, \bibinfo {author} {\bibfnamefont {U.}~\bibnamefont {Waghmare}},
  \bibinfo {author} {\bibfnamefont {X.}~\bibnamefont {Wu}}, \emph {et~al.},\
  }\bibfield  {title} {\bibinfo {title} {Accurate first-principles structures
  and energies of diversely bonded systems from an efficient density
  functional},\ }\href {https://doi.org/doi.org/10.1038/NCHEM.2535} {\bibfield
  {journal} {\bibinfo  {journal} {Nat. Chem.}\ }\textbf {\bibinfo {volume}
  {8}},\ \bibinfo {pages} {831} (\bibinfo {year}
  {2016}{\natexlab{a}})}\BibitemShut {NoStop}%
\bibitem [{\citenamefont {Kitchaev}\ \emph {et~al.}(2016)\citenamefont
  {Kitchaev}, \citenamefont {Peng}, \citenamefont {Liu}, \citenamefont {Sun},
  \citenamefont {Perdew},\ and\ \citenamefont {Ceder}}]{PhysRevB.93.045132}%
  \BibitemOpen
  \bibfield  {author} {\bibinfo {author} {\bibfnamefont {D.~A.}\ \bibnamefont
  {Kitchaev}}, \bibinfo {author} {\bibfnamefont {H.}~\bibnamefont {Peng}},
  \bibinfo {author} {\bibfnamefont {Y.}~\bibnamefont {Liu}}, \bibinfo {author}
  {\bibfnamefont {J.}~\bibnamefont {Sun}}, \bibinfo {author} {\bibfnamefont
  {J.~P.}\ \bibnamefont {Perdew}},\ and\ \bibinfo {author} {\bibfnamefont
  {G.}~\bibnamefont {Ceder}},\ }\bibfield  {title} {\bibinfo {title}
  {Energetics of ${\mathrm{mno}}_{2}$ polymorphs in density functional
  theory},\ }\href {https://doi.org/10.1103/PhysRevB.93.045132} {\bibfield
  {journal} {\bibinfo  {journal} {Phys. Rev. B}\ }\textbf {\bibinfo {volume}
  {93}},\ \bibinfo {pages} {045132} (\bibinfo {year} {2016})}\BibitemShut
  {NoStop}%
\bibitem [{\citenamefont {Zhang}\ \emph {et~al.}(2017)\citenamefont {Zhang},
  \citenamefont {Sun}, \citenamefont {Perdew},\ and\ \citenamefont
  {Wu}}]{PhysRevB.96.035143}%
  \BibitemOpen
  \bibfield  {author} {\bibinfo {author} {\bibfnamefont {Y.}~\bibnamefont
  {Zhang}}, \bibinfo {author} {\bibfnamefont {J.}~\bibnamefont {Sun}}, \bibinfo
  {author} {\bibfnamefont {J.~P.}\ \bibnamefont {Perdew}},\ and\ \bibinfo
  {author} {\bibfnamefont {X.}~\bibnamefont {Wu}},\ }\bibfield  {title}
  {\bibinfo {title} {Comparative first-principles studies of prototypical
  ferroelectric materials by lda, gga, and scan meta-gga},\ }\href
  {https://doi.org/10.1103/PhysRevB.96.035143} {\bibfield  {journal} {\bibinfo
  {journal} {Phys. Rev. B}\ }\textbf {\bibinfo {volume} {96}},\ \bibinfo
  {pages} {035143} (\bibinfo {year} {2017})}\BibitemShut {NoStop}%
\bibitem [{\citenamefont {Bart{\'o}k}\ and\ \citenamefont
  {Yates}(2019)}]{bartok2019regularized}%
  \BibitemOpen
  \bibfield  {author} {\bibinfo {author} {\bibfnamefont {A.~P.}\ \bibnamefont
  {Bart{\'o}k}}\ and\ \bibinfo {author} {\bibfnamefont {J.~R.}\ \bibnamefont
  {Yates}},\ }\bibfield  {title} {\bibinfo {title} {Regularized scan
  functional},\ }\href {https://doi.org/10.1063/1.5094646} {\bibfield
  {journal} {\bibinfo  {journal} {J. Chem. Phys.}\ }\textbf {\bibinfo {volume}
  {150}},\ \bibinfo {pages} {161101} (\bibinfo {year} {2019})}\BibitemShut
  {NoStop}%
\bibitem [{\citenamefont {Furness}\ \emph {et~al.}(2020)\citenamefont
  {Furness}, \citenamefont {Kaplan}, \citenamefont {Ning}, \citenamefont
  {Perdew},\ and\ \citenamefont {Sun}}]{furness2020accurate}%
  \BibitemOpen
  \bibfield  {author} {\bibinfo {author} {\bibfnamefont {J.~W.}\ \bibnamefont
  {Furness}}, \bibinfo {author} {\bibfnamefont {A.~D.}\ \bibnamefont {Kaplan}},
  \bibinfo {author} {\bibfnamefont {J.}~\bibnamefont {Ning}}, \bibinfo {author}
  {\bibfnamefont {J.~P.}\ \bibnamefont {Perdew}},\ and\ \bibinfo {author}
  {\bibfnamefont {J.}~\bibnamefont {Sun}},\ }\bibfield  {title} {\bibinfo
  {title} {Accurate and numerically efficient r2scan meta-generalized gradient
  approximation},\ }\href {https://doi.org/doi.org/10.1021/acs.jpclett.0c02405}
  {\bibfield  {journal} {\bibinfo  {journal} {J. Phys. Chem. Lett.}\ }\textbf
  {\bibinfo {volume} {11}},\ \bibinfo {pages} {8208} (\bibinfo {year}
  {2020})}\BibitemShut {NoStop}%
\bibitem [{\citenamefont {Shao}\ \emph {et~al.}(2023)\citenamefont {Shao},
  \citenamefont {Liu}, \citenamefont {Franchini}, \citenamefont {Xia},\ and\
  \citenamefont {He}}]{PhysRevB.108.024306}%
  \BibitemOpen
  \bibfield  {author} {\bibinfo {author} {\bibfnamefont {X.}~\bibnamefont
  {Shao}}, \bibinfo {author} {\bibfnamefont {P.}~\bibnamefont {Liu}}, \bibinfo
  {author} {\bibfnamefont {C.}~\bibnamefont {Franchini}}, \bibinfo {author}
  {\bibfnamefont {Y.}~\bibnamefont {Xia}},\ and\ \bibinfo {author}
  {\bibfnamefont {J.}~\bibnamefont {He}},\ }\bibfield  {title} {\bibinfo
  {title} {Assessing the performance of exchange-correlation functionals on
  lattice constants of binary solids at room temperature within the
  quasiharmonic approximation},\ }\href
  {https://doi.org/10.1103/PhysRevB.108.024306} {\bibfield  {journal} {\bibinfo
   {journal} {Phys. Rev. B}\ }\textbf {\bibinfo {volume} {108}},\ \bibinfo
  {pages} {024306} (\bibinfo {year} {2023})}\BibitemShut {NoStop}%
\bibitem [{\citenamefont {Jain}\ and\ \citenamefont {McGaughey}(2015)}]{RN24}%
  \BibitemOpen
  \bibfield  {author} {\bibinfo {author} {\bibfnamefont {A.}~\bibnamefont
  {Jain}}\ and\ \bibinfo {author} {\bibfnamefont {A.~J.~H.}\ \bibnamefont
  {McGaughey}},\ }\bibfield  {title} {\bibinfo {title} {Effect of
  exchange–correlation on first-principles-driven lattice thermal
  conductivity predictions of crystalline silicon},\ }\href
  {https://doi.org/10.1016/j.commatsci.2015.08.014} {\bibfield  {journal}
  {\bibinfo  {journal} {Comput. Mater. Sci.}\ }\textbf {\bibinfo {volume}
  {110}},\ \bibinfo {pages} {115} (\bibinfo {year} {2015})}\BibitemShut
  {NoStop}%
\bibitem [{\citenamefont {Perdew}\ \emph {et~al.}(1992)\citenamefont {Perdew},
  \citenamefont {Chevary}, \citenamefont {Vosko}, \citenamefont {Jackson},
  \citenamefont {Pederson}, \citenamefont {Singh},\ and\ \citenamefont
  {Fiolhais}}]{PhysRevB.46.6671}%
  \BibitemOpen
  \bibfield  {author} {\bibinfo {author} {\bibfnamefont {J.~P.}\ \bibnamefont
  {Perdew}}, \bibinfo {author} {\bibfnamefont {J.~A.}\ \bibnamefont {Chevary}},
  \bibinfo {author} {\bibfnamefont {S.~H.}\ \bibnamefont {Vosko}}, \bibinfo
  {author} {\bibfnamefont {K.~A.}\ \bibnamefont {Jackson}}, \bibinfo {author}
  {\bibfnamefont {M.~R.}\ \bibnamefont {Pederson}}, \bibinfo {author}
  {\bibfnamefont {D.~J.}\ \bibnamefont {Singh}},\ and\ \bibinfo {author}
  {\bibfnamefont {C.}~\bibnamefont {Fiolhais}},\ }\bibfield  {title} {\bibinfo
  {title} {Atoms, molecules, solids, and surfaces: Applications of the
  generalized gradient approximation for exchange and correlation},\ }\href
  {https://doi.org/10.1103/PhysRevB.46.6671} {\bibfield  {journal} {\bibinfo
  {journal} {Phys. Rev. B}\ }\textbf {\bibinfo {volume} {46}},\ \bibinfo
  {pages} {6671} (\bibinfo {year} {1992})}\BibitemShut {NoStop}%
\bibitem [{\citenamefont {Becke}(1988)}]{PhysRevA.38.3098}%
  \BibitemOpen
  \bibfield  {author} {\bibinfo {author} {\bibfnamefont {A.~D.}\ \bibnamefont
  {Becke}},\ }\bibfield  {title} {\bibinfo {title} {Density-functional
  exchange-energy approximation with correct asymptotic behavior},\ }\href
  {https://doi.org/10.1103/PhysRevA.38.3098} {\bibfield  {journal} {\bibinfo
  {journal} {Phys. Rev. A}\ }\textbf {\bibinfo {volume} {38}},\ \bibinfo
  {pages} {3098} (\bibinfo {year} {1988})}\BibitemShut {NoStop}%
\bibitem [{\citenamefont {Lee}\ \emph {et~al.}(1988)\citenamefont {Lee},
  \citenamefont {Yang},\ and\ \citenamefont {Parr}}]{PhysRevB.37.785}%
  \BibitemOpen
  \bibfield  {author} {\bibinfo {author} {\bibfnamefont {C.}~\bibnamefont
  {Lee}}, \bibinfo {author} {\bibfnamefont {W.}~\bibnamefont {Yang}},\ and\
  \bibinfo {author} {\bibfnamefont {R.~G.}\ \bibnamefont {Parr}},\ }\bibfield
  {title} {\bibinfo {title} {Development of the colle-salvetti
  correlation-energy formula into a functional of the electron density},\
  }\href {https://doi.org/10.1103/PhysRevB.37.785} {\bibfield  {journal}
  {\bibinfo  {journal} {Phys. Rev. B}\ }\textbf {\bibinfo {volume} {37}},\
  \bibinfo {pages} {785} (\bibinfo {year} {1988})}\BibitemShut {NoStop}%
\bibitem [{\citenamefont {Qin}\ \emph {et~al.}(2018)\citenamefont {Qin},
  \citenamefont {Qin}, \citenamefont {Wang},\ and\ \citenamefont {Hu}}]{RN25}%
  \BibitemOpen
  \bibfield  {author} {\bibinfo {author} {\bibfnamefont {G.}~\bibnamefont
  {Qin}}, \bibinfo {author} {\bibfnamefont {Z.}~\bibnamefont {Qin}}, \bibinfo
  {author} {\bibfnamefont {H.}~\bibnamefont {Wang}},\ and\ \bibinfo {author}
  {\bibfnamefont {M.}~\bibnamefont {Hu}},\ }\bibfield  {title} {\bibinfo
  {title} {On the diversity in the thermal transport properties of graphene: A
  first-principles-benchmark study testing different exchange-correlation
  functionals},\ }\href {https://doi.org/10.1016/j.commatsci.2018.05.007}
  {\bibfield  {journal} {\bibinfo  {journal} {Comput. Mater. Sci.}\ }\textbf
  {\bibinfo {volume} {151}},\ \bibinfo {pages} {153} (\bibinfo {year}
  {2018})}\BibitemShut {NoStop}%
\bibitem [{\citenamefont {Taheri}\ \emph {et~al.}(2018)\citenamefont {Taheri},
  \citenamefont {Da~Silva},\ and\ \citenamefont {Amon}}]{RN26}%
  \BibitemOpen
  \bibfield  {author} {\bibinfo {author} {\bibfnamefont {A.}~\bibnamefont
  {Taheri}}, \bibinfo {author} {\bibfnamefont {C.}~\bibnamefont {Da~Silva}},\
  and\ \bibinfo {author} {\bibfnamefont {C.~H.}\ \bibnamefont {Amon}},\
  }\bibfield  {title} {\bibinfo {title} {First-principles phonon thermal
  transport in graphene: Effects of exchange-correlation and type of
  pseudopotential},\ }\href {https://doi.org/10.1063/1.5027619} {\bibfield
  {journal} {\bibinfo  {journal} {J. Appl. Phys.}\ }\textbf {\bibinfo {volume}
  {123}} (\bibinfo {year} {2018})}\BibitemShut {NoStop}%
\bibitem [{\citenamefont {Han}\ and\ \citenamefont
  {Ruan}(2023)}]{PhysRevB.108.L121412}%
  \BibitemOpen
  \bibfield  {author} {\bibinfo {author} {\bibfnamefont {Z.}~\bibnamefont
  {Han}}\ and\ \bibinfo {author} {\bibfnamefont {X.}~\bibnamefont {Ruan}},\
  }\bibfield  {title} {\bibinfo {title} {Thermal conductivity of monolayer
  graphene: Convergent and lower than diamond},\ }\href
  {https://doi.org/10.1103/PhysRevB.108.L121412} {\bibfield  {journal}
  {\bibinfo  {journal} {Phys. Rev. B}\ }\textbf {\bibinfo {volume} {108}},\
  \bibinfo {pages} {L121412} (\bibinfo {year} {2023})}\BibitemShut {NoStop}%
\bibitem [{\citenamefont {Dongre}\ \emph {et~al.}(2022)\citenamefont {Dongre},
  \citenamefont {Carrete}, \citenamefont {Mingo},\ and\ \citenamefont
  {Madsen}}]{RN28}%
  \BibitemOpen
  \bibfield  {author} {\bibinfo {author} {\bibfnamefont {B.}~\bibnamefont
  {Dongre}}, \bibinfo {author} {\bibfnamefont {J.}~\bibnamefont {Carrete}},
  \bibinfo {author} {\bibfnamefont {N.}~\bibnamefont {Mingo}},\ and\ \bibinfo
  {author} {\bibfnamefont {G.~K.~H.}\ \bibnamefont {Madsen}},\ }\bibfield
  {title} {\bibinfo {title} {Thermal conductivity of group-iii phosphides: The
  special case of gap}\ }\href {https://doi.org/Phys. Rev. B 106, 205202}
  {Phys. Rev. B 106, 205202} (\bibinfo {year} {2022})\BibitemShut {NoStop}%
\bibitem [{\citenamefont {Xia}\ \emph {et~al.}(2020{\natexlab{b}})\citenamefont
  {Xia}, \citenamefont {Pal}, \citenamefont {He}, \citenamefont
  {Ozoli{\c{n}}{\v{s}}},\ and\ \citenamefont
  {Wolverton}}]{xia2020particlelike}%
  \BibitemOpen
  \bibfield  {author} {\bibinfo {author} {\bibfnamefont {Y.}~\bibnamefont
  {Xia}}, \bibinfo {author} {\bibfnamefont {K.}~\bibnamefont {Pal}}, \bibinfo
  {author} {\bibfnamefont {J.}~\bibnamefont {He}}, \bibinfo {author}
  {\bibfnamefont {V.}~\bibnamefont {Ozoli{\c{n}}{\v{s}}}},\ and\ \bibinfo
  {author} {\bibfnamefont {C.}~\bibnamefont {Wolverton}},\ }\bibfield  {title}
  {\bibinfo {title} {Particlelike phonon propagation dominates ultralow lattice
  thermal conductivity in crystalline tl3vse4},\ }\href
  {https://doi.org/doi.org/10.1103/PhysRevLett.124.065901} {\bibfield
  {journal} {\bibinfo  {journal} {Phys. Rev. Lett.}\ }\textbf {\bibinfo
  {volume} {124}},\ \bibinfo {pages} {065901} (\bibinfo {year}
  {2020}{\natexlab{b}})}\BibitemShut {NoStop}%
\bibitem [{\citenamefont {Jain}(2020)}]{jain2020multichannel}%
  \BibitemOpen
  \bibfield  {author} {\bibinfo {author} {\bibfnamefont {A.}~\bibnamefont
  {Jain}},\ }\bibfield  {title} {\bibinfo {title} {Multichannel thermal
  transport in crystalline tl 3 vse 4},\ }\href
  {https://doi.org/doi.org/10.1103/PhysRevB.102.201201} {\bibfield  {journal}
  {\bibinfo  {journal} {Phys. Rev. B}\ }\textbf {\bibinfo {volume} {102}},\
  \bibinfo {pages} {201201} (\bibinfo {year} {2020})}\BibitemShut {NoStop}%
\bibitem [{\citenamefont {Klime\ifmmode~\check{s}\else \v{s}\fi{}}\ \emph
  {et~al.}(2011)\citenamefont {Klime\ifmmode~\check{s}\else \v{s}\fi{}},
  \citenamefont {Bowler},\ and\ \citenamefont
  {Michaelides}}]{PhysRevB.83.195131}%
  \BibitemOpen
  \bibfield  {author} {\bibinfo {author} {\bibfnamefont {J.~c.~v.}\
  \bibnamefont {Klime\ifmmode~\check{s}\else \v{s}\fi{}}}, \bibinfo {author}
  {\bibfnamefont {D.~R.}\ \bibnamefont {Bowler}},\ and\ \bibinfo {author}
  {\bibfnamefont {A.}~\bibnamefont {Michaelides}},\ }\bibfield  {title}
  {\bibinfo {title} {Van der waals density functionals applied to solids},\
  }\href {https://doi.org/10.1103/PhysRevB.83.195131} {\bibfield  {journal}
  {\bibinfo  {journal} {Phys. Rev. B}\ }\textbf {\bibinfo {volume} {83}},\
  \bibinfo {pages} {195131} (\bibinfo {year} {2011})}\BibitemShut {NoStop}%
\bibitem [{\citenamefont {Bl\"ochl}(1994)}]{PAW1}%
  \BibitemOpen
  \bibfield  {author} {\bibinfo {author} {\bibfnamefont {P.~E.}\ \bibnamefont
  {Bl\"ochl}},\ }\bibfield  {title} {\bibinfo {title} {Projector augmented-wave
  method},\ }\href {https://doi.org/10.1103/PhysRevB.50.17953} {\bibfield
  {journal} {\bibinfo  {journal} {Phys. Rev. B}\ }\textbf {\bibinfo {volume}
  {50}},\ \bibinfo {pages} {17953} (\bibinfo {year} {1994})}\BibitemShut
  {NoStop}%
\bibitem [{\citenamefont {Kresse}\ and\ \citenamefont {Joubert}(1999)}]{PAW2}%
  \BibitemOpen
  \bibfield  {author} {\bibinfo {author} {\bibfnamefont {G.}~\bibnamefont
  {Kresse}}\ and\ \bibinfo {author} {\bibfnamefont {D.}~\bibnamefont
  {Joubert}},\ }\bibfield  {title} {\bibinfo {title} {From ultrasoft
  pseudopotentials to the projector augmented-wave method},\ }\href
  {https://doi.org/10.1103/PhysRevB.59.1758} {\bibfield  {journal} {\bibinfo
  {journal} {Phys. Rev. B}\ }\textbf {\bibinfo {volume} {59}},\ \bibinfo
  {pages} {1758} (\bibinfo {year} {1999})}\BibitemShut {NoStop}%
\bibitem [{\citenamefont {Kresse}\ and\ \citenamefont
  {Furthm\"uller}(1996)}]{vasp1}%
  \BibitemOpen
  \bibfield  {author} {\bibinfo {author} {\bibfnamefont {G.}~\bibnamefont
  {Kresse}}\ and\ \bibinfo {author} {\bibfnamefont {J.}~\bibnamefont
  {Furthm\"uller}},\ }\bibfield  {title} {\bibinfo {title} {{Efficient
  iterative schemes for {\sl ab-initio} total-energy calculations using a
  plane-wave basis set}},\ }\href {https://doi.org/10.1103/PhysRevB.54.11169}
  {\bibfield  {journal} {\bibinfo  {journal} {Phys. Rev. B}\ }\textbf {\bibinfo
  {volume} {54}},\ \bibinfo {pages} {11169} (\bibinfo {year}
  {1996})}\BibitemShut {NoStop}%
\bibitem [{\citenamefont {Kresse}\ and\ \citenamefont
  {Furthm\"{u}ller}(1996)}]{vasp2}%
  \BibitemOpen
  \bibfield  {author} {\bibinfo {author} {\bibfnamefont {G.}~\bibnamefont
  {Kresse}}\ and\ \bibinfo {author} {\bibfnamefont {J.}~\bibnamefont
  {Furthm\"{u}ller}},\ }\bibfield  {title} {\bibinfo {title} {{Efficiency of
  ab-initio total energy calculations for metals and semiconductors using a
  plane-wave basis set}},\ }\href
  {https://doi.org/10.1016/0927-0256(96)00008-0} {\bibfield  {journal}
  {\bibinfo  {journal} {Comput. Mater. Sci.}\ }\textbf {\bibinfo {volume}
  {6}},\ \bibinfo {pages} {15} (\bibinfo {year} {1996})}\BibitemShut {NoStop}%
\bibitem [{\citenamefont {Sun}\ \emph {et~al.}(2016{\natexlab{b}})\citenamefont
  {Sun}, \citenamefont {Remsing}, \citenamefont {Zhang}, \citenamefont {Sun},
  \citenamefont {Ruzsinszky}, \citenamefont {Peng}, \citenamefont {Yang},
  \citenamefont {Paul}, \citenamefont {Waghmare}, \citenamefont {Wu},
  \citenamefont {Klein},\ and\ \citenamefont {Perdew}}]{RN52}%
  \BibitemOpen
  \bibfield  {author} {\bibinfo {author} {\bibfnamefont {J.}~\bibnamefont
  {Sun}}, \bibinfo {author} {\bibfnamefont {R.~C.}\ \bibnamefont {Remsing}},
  \bibinfo {author} {\bibfnamefont {Y.}~\bibnamefont {Zhang}}, \bibinfo
  {author} {\bibfnamefont {Z.}~\bibnamefont {Sun}}, \bibinfo {author}
  {\bibfnamefont {A.}~\bibnamefont {Ruzsinszky}}, \bibinfo {author}
  {\bibfnamefont {H.}~\bibnamefont {Peng}}, \bibinfo {author} {\bibfnamefont
  {Z.}~\bibnamefont {Yang}}, \bibinfo {author} {\bibfnamefont {A.}~\bibnamefont
  {Paul}}, \bibinfo {author} {\bibfnamefont {U.}~\bibnamefont {Waghmare}},
  \bibinfo {author} {\bibfnamefont {X.}~\bibnamefont {Wu}}, \bibinfo {author}
  {\bibfnamefont {M.~L.}\ \bibnamefont {Klein}},\ and\ \bibinfo {author}
  {\bibfnamefont {J.~P.}\ \bibnamefont {Perdew}},\ }\bibfield  {title}
  {\bibinfo {title} {Accurate first-principles structures and energies of
  diversely bonded systems from an efficient density functional},\ }\href
  {https://doi.org/10.1038/nchem.2535} {\bibfield  {journal} {\bibinfo
  {journal} {Nat Chem}\ }\textbf {\bibinfo {volume} {8}},\ \bibinfo {pages}
  {831} (\bibinfo {year} {2016}{\natexlab{b}})}\BibitemShut {NoStop}%
\bibitem [{\citenamefont {Fu}\ and\ \citenamefont {Singh}(2018)}]{RN53}%
  \BibitemOpen
  \bibfield  {author} {\bibinfo {author} {\bibfnamefont {Y.}~\bibnamefont
  {Fu}}\ and\ \bibinfo {author} {\bibfnamefont {D.~J.}\ \bibnamefont {Singh}},\
  }\bibfield  {title} {\bibinfo {title} {Applicability of the strongly
  constrained and appropriately normed density functional to transition-metal
  magnetism},\ }\href {https://doi.org/10.1103/PhysRevLett.121.207201}
  {\bibfield  {journal} {\bibinfo  {journal} {Phys Rev Lett}\ }\textbf
  {\bibinfo {volume} {121}},\ \bibinfo {pages} {207201} (\bibinfo {year}
  {2018})}\BibitemShut {NoStop}%
\bibitem [{\citenamefont {Mejia-Rodriguez}\ and\ \citenamefont
  {Trickey}(2019)}]{RN54}%
  \BibitemOpen
  \bibfield  {author} {\bibinfo {author} {\bibfnamefont {D.}~\bibnamefont
  {Mejia-Rodriguez}}\ and\ \bibinfo {author} {\bibfnamefont {S.~B.}\
  \bibnamefont {Trickey}},\ }\bibfield  {title} {\bibinfo {title} {Comment on
  "regularized scan functional" [j. chem. phys. 150, 161101 (2019)]},\ }\href
  {https://doi.org/10.1063/1.5120408} {\bibfield  {journal} {\bibinfo
  {journal} {J Chem Phys}\ }\textbf {\bibinfo {volume} {151}},\ \bibinfo
  {pages} {207101} (\bibinfo {year} {2019})}\BibitemShut {NoStop}%
\bibitem [{\citenamefont {Togo}\ and\ \citenamefont
  {Tanaka}(2015)}]{TOGO20151}%
  \BibitemOpen
  \bibfield  {author} {\bibinfo {author} {\bibfnamefont {A.}~\bibnamefont
  {Togo}}\ and\ \bibinfo {author} {\bibfnamefont {I.}~\bibnamefont {Tanaka}},\
  }\bibfield  {title} {\bibinfo {title} {First principles phonon calculations
  in materials science},\ }\href
  {https://doi.org/https://doi.org/10.1016/j.scriptamat.2015.07.021} {\bibfield
   {journal} {\bibinfo  {journal} {Scr. Mater.}\ }\textbf {\bibinfo {volume}
  {108}},\ \bibinfo {pages} {1} (\bibinfo {year} {2015})}\BibitemShut {NoStop}%
\bibitem [{\citenamefont {Gonze}\ \emph {et~al.}(1994)\citenamefont {Gonze},
  \citenamefont {Charlier}, \citenamefont {Allan},\ and\ \citenamefont
  {Teter}}]{PhysRevB.50.13035}%
  \BibitemOpen
  \bibfield  {author} {\bibinfo {author} {\bibfnamefont {X.}~\bibnamefont
  {Gonze}}, \bibinfo {author} {\bibfnamefont {J.-C.}\ \bibnamefont {Charlier}},
  \bibinfo {author} {\bibfnamefont {D.}~\bibnamefont {Allan}},\ and\ \bibinfo
  {author} {\bibfnamefont {M.}~\bibnamefont {Teter}},\ }\bibfield  {title}
  {\bibinfo {title} {Interatomic force constants from first principles: The
  case of \ensuremath{\alpha}-quartz},\ }\href
  {https://doi.org/10.1103/PhysRevB.50.13035} {\bibfield  {journal} {\bibinfo
  {journal} {Phys. Rev. B}\ }\textbf {\bibinfo {volume} {50}},\ \bibinfo
  {pages} {13035} (\bibinfo {year} {1994})}\BibitemShut {NoStop}%
\bibitem [{\citenamefont {Gonze}\ and\ \citenamefont
  {Lee}(1997)}]{PhysRevB.55.10355}%
  \BibitemOpen
  \bibfield  {author} {\bibinfo {author} {\bibfnamefont {X.}~\bibnamefont
  {Gonze}}\ and\ \bibinfo {author} {\bibfnamefont {C.}~\bibnamefont {Lee}},\
  }\bibfield  {title} {\bibinfo {title} {Dynamical matrices, born effective
  charges, dielectric permittivity tensors, and interatomic force constants
  from density-functional perturbation theory},\ }\href
  {https://doi.org/10.1103/PhysRevB.55.10355} {\bibfield  {journal} {\bibinfo
  {journal} {Phys. Rev. B}\ }\textbf {\bibinfo {volume} {55}},\ \bibinfo
  {pages} {10355} (\bibinfo {year} {1997})}\BibitemShut {NoStop}%
\bibitem [{\citenamefont {Zhou}\ \emph {et~al.}(2014)\citenamefont {Zhou},
  \citenamefont {Nielson}, \citenamefont {Xia},\ and\ \citenamefont
  {Ozolins}}]{RN38}%
  \BibitemOpen
  \bibfield  {author} {\bibinfo {author} {\bibfnamefont {F.}~\bibnamefont
  {Zhou}}, \bibinfo {author} {\bibfnamefont {W.}~\bibnamefont {Nielson}},
  \bibinfo {author} {\bibfnamefont {Y.}~\bibnamefont {Xia}},\ and\ \bibinfo
  {author} {\bibfnamefont {V.}~\bibnamefont {Ozolins}},\ }\bibfield  {title}
  {\bibinfo {title} {Lattice anharmonicity and thermal conductivity from
  compressive sensing of first-principles calculations},\ }\href
  {https://doi.org/10.1103/PhysRevLett.113.185501} {\bibfield  {journal}
  {\bibinfo  {journal} {Phys Rev Lett}\ }\textbf {\bibinfo {volume} {113}},\
  \bibinfo {pages} {185501} (\bibinfo {year} {2014})}\BibitemShut {NoStop}%
\bibitem [{\citenamefont {Hooton}(1955)}]{doi:10.1080/14786440408520575}%
  \BibitemOpen
  \bibfield  {author} {\bibinfo {author} {\bibfnamefont {D.}~\bibnamefont
  {Hooton}},\ }\bibfield  {title} {\bibinfo {title} {Li. a new treatment of
  anharmonicity in lattice thermodynamics: I},\ }\href
  {https://doi.org/10.1080/14786440408520575} {\bibfield  {journal} {\bibinfo
  {journal} {Philos. Mag. J. Sci.}\ }\textbf {\bibinfo {volume} {46}},\
  \bibinfo {pages} {422} (\bibinfo {year} {1955})}\BibitemShut {NoStop}%
\bibitem [{\citenamefont {Pawley}\ \emph {et~al.}(1966)\citenamefont {Pawley},
  \citenamefont {Cochran}, \citenamefont {Cowley},\ and\ \citenamefont
  {Dolling}}]{PhysRevLett.17.753}%
  \BibitemOpen
  \bibfield  {author} {\bibinfo {author} {\bibfnamefont {G.~S.}\ \bibnamefont
  {Pawley}}, \bibinfo {author} {\bibfnamefont {W.}~\bibnamefont {Cochran}},
  \bibinfo {author} {\bibfnamefont {R.~A.}\ \bibnamefont {Cowley}},\ and\
  \bibinfo {author} {\bibfnamefont {G.}~\bibnamefont {Dolling}},\ }\bibfield
  {title} {\bibinfo {title} {Diatomic ferroelectrics},\ }\href
  {https://doi.org/10.1103/PhysRevLett.17.753} {\bibfield  {journal} {\bibinfo
  {journal} {Phys. Rev. Lett.}\ }\textbf {\bibinfo {volume} {17}},\ \bibinfo
  {pages} {753} (\bibinfo {year} {1966})}\BibitemShut {NoStop}%
\bibitem [{\citenamefont {Werthamer}(1970)}]{PhysRevB.1.572}%
  \BibitemOpen
  \bibfield  {author} {\bibinfo {author} {\bibfnamefont {N.~R.}\ \bibnamefont
  {Werthamer}},\ }\bibfield  {title} {\bibinfo {title} {Self-consistent phonon
  formulation of anharmonic lattice dynamics},\ }\href
  {https://doi.org/10.1103/PhysRevB.1.572} {\bibfield  {journal} {\bibinfo
  {journal} {Phys. Rev. B}\ }\textbf {\bibinfo {volume} {1}},\ \bibinfo {pages}
  {572} (\bibinfo {year} {1970})}\BibitemShut {NoStop}%
\bibitem [{\citenamefont {Guo}\ \emph {et~al.}(2024)\citenamefont {Guo},
  \citenamefont {Han}, \citenamefont {Feng}, \citenamefont {Lin},\ and\
  \citenamefont {Ruan}}]{guo2024sampling}%
  \BibitemOpen
  \bibfield  {author} {\bibinfo {author} {\bibfnamefont {Z.}~\bibnamefont
  {Guo}}, \bibinfo {author} {\bibfnamefont {Z.}~\bibnamefont {Han}}, \bibinfo
  {author} {\bibfnamefont {D.}~\bibnamefont {Feng}}, \bibinfo {author}
  {\bibfnamefont {G.}~\bibnamefont {Lin}},\ and\ \bibinfo {author}
  {\bibfnamefont {X.}~\bibnamefont {Ruan}},\ }\bibfield  {title} {\bibinfo
  {title} {Sampling-accelerated prediction of phonon scattering rates for
  converged thermal conductivity and radiative properties},\ }\href
  {https://doi.org/https://www.nature.com/articles/s41524-024-01215-8}
  {\bibfield  {journal} {\bibinfo  {journal} {npj Comput. Mater.}\ }\textbf
  {\bibinfo {volume} {10}},\ \bibinfo {pages} {31} (\bibinfo {year}
  {2024})}\BibitemShut {NoStop}%
\bibitem [{\citenamefont {Ong}\ \emph {et~al.}(2013)\citenamefont {Ong},
  \citenamefont {Richards}, \citenamefont {Jain}, \citenamefont {Hautier},
  \citenamefont {Kocher}, \citenamefont {Cholia}, \citenamefont {Gunter},
  \citenamefont {Chevrier}, \citenamefont {Persson},\ and\ \citenamefont
  {Ceder}}]{ONG2013314}%
  \BibitemOpen
  \bibfield  {author} {\bibinfo {author} {\bibfnamefont {S.~P.}\ \bibnamefont
  {Ong}}, \bibinfo {author} {\bibfnamefont {W.~D.}\ \bibnamefont {Richards}},
  \bibinfo {author} {\bibfnamefont {A.}~\bibnamefont {Jain}}, \bibinfo {author}
  {\bibfnamefont {G.}~\bibnamefont {Hautier}}, \bibinfo {author} {\bibfnamefont
  {M.}~\bibnamefont {Kocher}}, \bibinfo {author} {\bibfnamefont
  {S.}~\bibnamefont {Cholia}}, \bibinfo {author} {\bibfnamefont
  {D.}~\bibnamefont {Gunter}}, \bibinfo {author} {\bibfnamefont {V.~L.}\
  \bibnamefont {Chevrier}}, \bibinfo {author} {\bibfnamefont {K.~A.}\
  \bibnamefont {Persson}},\ and\ \bibinfo {author} {\bibfnamefont
  {G.}~\bibnamefont {Ceder}},\ }\bibfield  {title} {\bibinfo {title} {Python
  materials genomics (pymatgen): A robust, open-source python library for
  materials analysis},\ }\href
  {https://doi.org/https://doi.org/10.1016/j.commatsci.2012.10.028} {\bibfield
  {journal} {\bibinfo  {journal} {Comput. Mater. Sci.}\ }\textbf {\bibinfo
  {volume} {68}},\ \bibinfo {pages} {314} (\bibinfo {year} {2013})}\BibitemShut
  {NoStop}%
\bibitem [{\citenamefont {Hill}(1952)}]{R-Hill_1952}%
  \BibitemOpen
  \bibfield  {author} {\bibinfo {author} {\bibfnamefont {R.}~\bibnamefont
  {Hill}},\ }\bibfield  {title} {\bibinfo {title} {The elastic behaviour of a
  crystalline aggregate},\ }\href {https://doi.org/10.1088/0370-1298/65/5/307}
  {\bibfield  {journal} {\bibinfo  {journal} {Proc. Phys. Soc. A}\ }\textbf
  {\bibinfo {volume} {65}},\ \bibinfo {pages} {349} (\bibinfo {year}
  {1952})}\BibitemShut {NoStop}%
\bibitem [{\citenamefont {Tritt}(2010)}]{tritt2005thermal}%
  \BibitemOpen
  \bibfield  {author} {\bibinfo {author} {\bibfnamefont {T.~M.}\ \bibnamefont
  {Tritt}},\ }\href {https://link.springer.com/book/10.1007/b136496} {\emph
  {\bibinfo {title} {Thermal Conductivity: Theory, Properties, and
  Applications}}}\ (\bibinfo {year} {2010})\BibitemShut {NoStop}%
\bibitem [{\citenamefont {Koehler}(1966)}]{PhysRevLett.17.89}%
  \BibitemOpen
  \bibfield  {author} {\bibinfo {author} {\bibfnamefont {T.~R.}\ \bibnamefont
  {Koehler}},\ }\bibfield  {title} {\bibinfo {title} {Theory of the
  self-consistent harmonic approximation with application to solid neon},\
  }\href {https://doi.org/10.1103/PhysRevLett.17.89} {\bibfield  {journal}
  {\bibinfo  {journal} {Phys. Rev. Lett.}\ }\textbf {\bibinfo {volume} {17}},\
  \bibinfo {pages} {89} (\bibinfo {year} {1966})}\BibitemShut {NoStop}%
\bibitem [{\citenamefont {Xia}\ \emph {et~al.}(2020{\natexlab{c}})\citenamefont
  {Xia}, \citenamefont {Ozoli\ifmmode \mbox{\c{n}}\else
  \c{n}\fi{}\ifmmode~\check{s}\else \v{s}\fi{}},\ and\ \citenamefont
  {Wolverton}}]{PhysRevLett.125.085901}%
  \BibitemOpen
  \bibfield  {author} {\bibinfo {author} {\bibfnamefont {Y.}~\bibnamefont
  {Xia}}, \bibinfo {author} {\bibfnamefont {V.}~\bibnamefont {Ozoli\ifmmode
  \mbox{\c{n}}\else \c{n}\fi{}\ifmmode~\check{s}\else \v{s}\fi{}}},\ and\
  \bibinfo {author} {\bibfnamefont {C.}~\bibnamefont {Wolverton}},\ }\bibfield
  {title} {\bibinfo {title} {Microscopic mechanisms of glasslike lattice
  thermal transport in cubic
  ${\mathrm{cu}}_{12}{\mathrm{sb}}_{4}{\mathrm{s}}_{13}$ tetrahedrites},\
  }\href {https://doi.org/10.1103/PhysRevLett.125.085901} {\bibfield  {journal}
  {\bibinfo  {journal} {Phys. Rev. Lett.}\ }\textbf {\bibinfo {volume} {125}},\
  \bibinfo {pages} {085901} (\bibinfo {year} {2020}{\natexlab{c}})}\BibitemShut
  {NoStop}%
\bibitem [{\citenamefont {Hanus}\ \emph {et~al.}(2019)\citenamefont {Hanus},
  \citenamefont {Agne}, \citenamefont {Rettie}, \citenamefont {Chen},
  \citenamefont {Tan}, \citenamefont {Chung}, \citenamefont {Kanatzidis},
  \citenamefont {Pei}, \citenamefont {Voorhees},\ and\ \citenamefont
  {Snyder}}]{https://doi.org/10.1002/adma.201900108}%
  \BibitemOpen
  \bibfield  {author} {\bibinfo {author} {\bibfnamefont {R.}~\bibnamefont
  {Hanus}}, \bibinfo {author} {\bibfnamefont {M.~T.}\ \bibnamefont {Agne}},
  \bibinfo {author} {\bibfnamefont {A.~J.~E.}\ \bibnamefont {Rettie}}, \bibinfo
  {author} {\bibfnamefont {Z.}~\bibnamefont {Chen}}, \bibinfo {author}
  {\bibfnamefont {G.}~\bibnamefont {Tan}}, \bibinfo {author} {\bibfnamefont
  {D.~Y.}\ \bibnamefont {Chung}}, \bibinfo {author} {\bibfnamefont {M.~G.}\
  \bibnamefont {Kanatzidis}}, \bibinfo {author} {\bibfnamefont
  {Y.}~\bibnamefont {Pei}}, \bibinfo {author} {\bibfnamefont {P.~W.}\
  \bibnamefont {Voorhees}},\ and\ \bibinfo {author} {\bibfnamefont {G.~J.}\
  \bibnamefont {Snyder}},\ }\bibfield  {title} {\bibinfo {title} {Lattice
  softening significantly reduces thermal conductivity and leads to high
  thermoelectric efficiency},\ }\href
  {https://doi.org/https://doi.org/10.1002/adma.201900108} {\bibfield
  {journal} {\bibinfo  {journal} {Adv. Mater.}\ }\textbf {\bibinfo {volume}
  {31}},\ \bibinfo {pages} {1900108} (\bibinfo {year} {2019})}\BibitemShut
  {NoStop}%
\bibitem [{\citenamefont {Chen}\ \emph {et~al.}(2018)\citenamefont {Chen},
  \citenamefont {Zhang}, \citenamefont {Lin}, \citenamefont {Chen},\ and\
  \citenamefont {Pei}}]{chen2018rationalizing}%
  \BibitemOpen
  \bibfield  {author} {\bibinfo {author} {\bibfnamefont {Z.}~\bibnamefont
  {Chen}}, \bibinfo {author} {\bibfnamefont {X.}~\bibnamefont {Zhang}},
  \bibinfo {author} {\bibfnamefont {S.}~\bibnamefont {Lin}}, \bibinfo {author}
  {\bibfnamefont {L.}~\bibnamefont {Chen}},\ and\ \bibinfo {author}
  {\bibfnamefont {Y.}~\bibnamefont {Pei}},\ }\bibfield  {title} {\bibinfo
  {title} {Rationalizing phonon dispersion for lattice thermal conductivity of
  solids},\ }\href {https://doi.org/10.1093/nsr/nwy097} {\bibfield  {journal}
  {\bibinfo  {journal} {Natl. Sci. Rev.}\ }\textbf {\bibinfo {volume} {5}},\
  \bibinfo {pages} {888} (\bibinfo {year} {2018})}\BibitemShut {NoStop}%
\bibitem [{\citenamefont
  {Grüneisen}(1908)}]{https://doi.org/10.1002/andp.19083310611}%
  \BibitemOpen
  \bibfield  {author} {\bibinfo {author} {\bibfnamefont {E.}~\bibnamefont
  {Grüneisen}},\ }\bibfield  {title} {\bibinfo {title} {Über die thermische
  ausdehnung und die spezifische wärme der metalle},\ }\href
  {https://doi.org/10.1002/andp.19083310611} {\bibfield  {journal} {\bibinfo
  {journal} {Ann. Phys.}\ }\textbf {\bibinfo {volume} {331}},\ \bibinfo {pages}
  {211} (\bibinfo {year} {1908})}\BibitemShut {NoStop}%
\bibitem [{\citenamefont
  {Grüneisen}(1912)}]{https://doi.org/10.1002/andp.19123441202}%
  \BibitemOpen
  \bibfield  {author} {\bibinfo {author} {\bibfnamefont {E.}~\bibnamefont
  {Grüneisen}},\ }\bibfield  {title} {\bibinfo {title} {Theorie des festen
  zustandes einatomiger elemente},\ }\href
  {https://doi.org/https://doi.org/10.1002/andp.19123441202} {\bibfield
  {journal} {\bibinfo  {journal} {Ann. Phys.}\ }\textbf {\bibinfo {volume}
  {344}},\ \bibinfo {pages} {257} (\bibinfo {year} {1912})}\BibitemShut
  {NoStop}%
\bibitem [{\citenamefont {Morelli}\ and\ \citenamefont
  {Slack}(2006)}]{morelli2006high}%
  \BibitemOpen
  \bibfield  {author} {\bibinfo {author} {\bibfnamefont {D.~T.}\ \bibnamefont
  {Morelli}}\ and\ \bibinfo {author} {\bibfnamefont {G.~A.}\ \bibnamefont
  {Slack}},\ }\bibfield  {title} {\bibinfo {title} {High lattice thermal
  conductivity solids},\ }in\ \href {https://doi.org/10.1007/0-387-25100-6_2}
  {\emph {\bibinfo {booktitle} {High thermal conductivity materials}}}\
  (\bibinfo  {publisher} {Springer},\ \bibinfo {year} {2006})\ pp.\ \bibinfo
  {pages} {37--68}\BibitemShut {NoStop}%
\bibitem [{\citenamefont {Ritz}\ \emph {et~al.}(2019)\citenamefont {Ritz},
  \citenamefont {Li},\ and\ \citenamefont {Benedek}}]{ritz2019thermal}%
  \BibitemOpen
  \bibfield  {author} {\bibinfo {author} {\bibfnamefont {E.~T.}\ \bibnamefont
  {Ritz}}, \bibinfo {author} {\bibfnamefont {S.~J.}\ \bibnamefont {Li}},\ and\
  \bibinfo {author} {\bibfnamefont {N.~A.}\ \bibnamefont {Benedek}},\
  }\bibfield  {title} {\bibinfo {title} {Thermal expansion in insulating solids
  from first principles},\ }\href {https://doi.org/10.1063/1.5125779}
  {\bibfield  {journal} {\bibinfo  {journal} {J. Appl. Phys.}\ }\textbf
  {\bibinfo {volume} {126}} (\bibinfo {year} {2019})}\BibitemShut {NoStop}%
\bibitem [{\citenamefont {Li}\ \emph {et~al.}(2014)\citenamefont {Li},
  \citenamefont {Carrete}, \citenamefont {Katcho},\ and\ \citenamefont
  {Mingo}}]{ShengBTE_2014}%
  \BibitemOpen
  \bibfield  {author} {\bibinfo {author} {\bibfnamefont {W.}~\bibnamefont
  {Li}}, \bibinfo {author} {\bibfnamefont {J.}~\bibnamefont {Carrete}},
  \bibinfo {author} {\bibfnamefont {N.~A.}\ \bibnamefont {Katcho}},\ and\
  \bibinfo {author} {\bibfnamefont {N.}~\bibnamefont {Mingo}},\ }\bibfield
  {title} {\bibinfo {title} {{ShengBTE:} a solver of the {B}oltzmann transport
  equation for phonons},\ }\href {https://doi.org/10.1016/j.cpc.2014.02.015}
  {\bibfield  {journal} {\bibinfo  {journal} {Comp. Phys. Commun.}\ }\textbf
  {\bibinfo {volume} {185}},\ \bibinfo {pages} {1747–1758} (\bibinfo {year}
  {2014})}\BibitemShut {NoStop}%
\bibitem [{\citenamefont {Shind{\'e}}\ and\ \citenamefont
  {Goela}(2006)}]{shinde2006high}%
  \BibitemOpen
  \bibfield  {author} {\bibinfo {author} {\bibfnamefont {S.~L.}\ \bibnamefont
  {Shind{\'e}}}\ and\ \bibinfo {author} {\bibfnamefont {J.}~\bibnamefont
  {Goela}},\ }\href {https://link.springer.com/book/10.1007/b106785} {\emph
  {\bibinfo {title} {High thermal conductivity materials}}},\ Vol.~\bibinfo
  {volume} {91}\ (\bibinfo  {publisher} {Springer},\ \bibinfo {year}
  {2006})\BibitemShut {NoStop}%
\bibitem [{\citenamefont {Slack}(1979)}]{slack1979thermal}%
  \BibitemOpen
  \bibfield  {author} {\bibinfo {author} {\bibfnamefont {G.~A.}\ \bibnamefont
  {Slack}},\ }\bibfield  {title} {\bibinfo {title} {The thermal conductivity of
  nonmetallic crystals},\ }\href
  {https://doi.org/10.1016/S0081-1947(08)60359-8} {\bibfield  {journal}
  {\bibinfo  {journal} {SOLID STATE PHYS}\ }\textbf {\bibinfo {volume} {34}},\
  \bibinfo {pages} {1} (\bibinfo {year} {1979})}\BibitemShut {NoStop}%
\bibitem [{\citenamefont {Miller}\ \emph {et~al.}(2017)\citenamefont {Miller},
  \citenamefont {Gorai}, \citenamefont {Ortiz}, \citenamefont {Goyal},
  \citenamefont {Gao}, \citenamefont {Barnett}, \citenamefont {Mason},
  \citenamefont {Snyder}, \citenamefont {Lv}, \citenamefont {Stevanović},\
  and\ \citenamefont {Toberer}}]{doi:10.1021/acs.chemmater.6b04179}%
  \BibitemOpen
  \bibfield  {author} {\bibinfo {author} {\bibfnamefont {S.~A.}\ \bibnamefont
  {Miller}}, \bibinfo {author} {\bibfnamefont {P.}~\bibnamefont {Gorai}},
  \bibinfo {author} {\bibfnamefont {B.~R.}\ \bibnamefont {Ortiz}}, \bibinfo
  {author} {\bibfnamefont {A.}~\bibnamefont {Goyal}}, \bibinfo {author}
  {\bibfnamefont {D.}~\bibnamefont {Gao}}, \bibinfo {author} {\bibfnamefont
  {S.~A.}\ \bibnamefont {Barnett}}, \bibinfo {author} {\bibfnamefont {T.~O.}\
  \bibnamefont {Mason}}, \bibinfo {author} {\bibfnamefont {G.~J.}\ \bibnamefont
  {Snyder}}, \bibinfo {author} {\bibfnamefont {Q.}~\bibnamefont {Lv}}, \bibinfo
  {author} {\bibfnamefont {V.}~\bibnamefont {Stevanović}},\ and\ \bibinfo
  {author} {\bibfnamefont {E.~S.}\ \bibnamefont {Toberer}},\ }\bibfield
  {title} {\bibinfo {title} {Capturing anharmonicity in a lattice thermal
  conductivity model for high-throughput predictions},\ }\href
  {https://doi.org/10.1021/acs.chemmater.6b04179} {\bibfield  {journal}
  {\bibinfo  {journal} {Chem. Mater.}\ }\textbf {\bibinfo {volume} {29}},\
  \bibinfo {pages} {2494} (\bibinfo {year} {2017})}\BibitemShut {NoStop}%
\bibitem [{\citenamefont {Peierls}(1996)}]{peierls1996quantum}%
  \BibitemOpen
  \bibfield  {author} {\bibinfo {author} {\bibfnamefont {R.~E.}\ \bibnamefont
  {Peierls}},\ }\href@noop {} {\emph {\bibinfo {title} {Quantum theory of
  solids}}}\ (\bibinfo  {publisher} {Clarendon Press},\ \bibinfo {year}
  {1996})\BibitemShut {NoStop}%
\bibitem [{\citenamefont {Xie}\ \emph {et~al.}(2020)\citenamefont {Xie},
  \citenamefont {Feng}, \citenamefont {Li},\ and\ \citenamefont
  {He}}]{PhysRevLett.125.245901}%
  \BibitemOpen
  \bibfield  {author} {\bibinfo {author} {\bibfnamefont {L.}~\bibnamefont
  {Xie}}, \bibinfo {author} {\bibfnamefont {J.~H.}\ \bibnamefont {Feng}},
  \bibinfo {author} {\bibfnamefont {R.}~\bibnamefont {Li}},\ and\ \bibinfo
  {author} {\bibfnamefont {J.~Q.}\ \bibnamefont {He}},\ }\bibfield  {title}
  {\bibinfo {title} {First-principles study of anharmonic lattice dynamics in
  low thermal conductivity ${\mathrm{agcrse}}_{2}$: Evidence for a large
  resonant four-phonon scattering},\ }\href
  {https://doi.org/10.1103/PhysRevLett.125.245901} {\bibfield  {journal}
  {\bibinfo  {journal} {Phys. Rev. Lett.}\ }\textbf {\bibinfo {volume} {125}},\
  \bibinfo {pages} {245901} (\bibinfo {year} {2020})}\BibitemShut {NoStop}%
\bibitem [{\citenamefont {Feng}\ and\ \citenamefont
  {Ruan}(2016)}]{PhysRevB.93.045202}%
  \BibitemOpen
  \bibfield  {author} {\bibinfo {author} {\bibfnamefont {T.}~\bibnamefont
  {Feng}}\ and\ \bibinfo {author} {\bibfnamefont {X.}~\bibnamefont {Ruan}},\
  }\bibfield  {title} {\bibinfo {title} {Quantum mechanical prediction of
  four-phonon scattering rates and reduced thermal conductivity of solids},\
  }\href {https://link.aps.org/doi/10.1103/PhysRevB.93.045202} {\bibfield
  {journal} {\bibinfo  {journal} {Phys. Rev. B}\ }\textbf {\bibinfo {volume}
  {93}},\ \bibinfo {pages} {045202} (\bibinfo {year} {2016})}\BibitemShut
  {NoStop}%
\bibitem [{\citenamefont {Zeier}\ \emph {et~al.}(2016)\citenamefont {Zeier},
  \citenamefont {Zevalkink}, \citenamefont {Gibbs}, \citenamefont {Hautier},
  \citenamefont {Kanatzidis},\ and\ \citenamefont
  {Snyder}}]{https://doi.org/10.1002/anie.201508381}%
  \BibitemOpen
  \bibfield  {author} {\bibinfo {author} {\bibfnamefont {W.~G.}\ \bibnamefont
  {Zeier}}, \bibinfo {author} {\bibfnamefont {A.}~\bibnamefont {Zevalkink}},
  \bibinfo {author} {\bibfnamefont {Z.~M.}\ \bibnamefont {Gibbs}}, \bibinfo
  {author} {\bibfnamefont {G.}~\bibnamefont {Hautier}}, \bibinfo {author}
  {\bibfnamefont {M.~G.}\ \bibnamefont {Kanatzidis}},\ and\ \bibinfo {author}
  {\bibfnamefont {G.~J.}\ \bibnamefont {Snyder}},\ }\bibfield  {title}
  {\bibinfo {title} {Thinking like a chemist: Intuition in thermoelectric
  materials},\ }\href {https://doi.org/https://doi.org/10.1002/anie.201508381}
  {\bibfield  {journal} {\bibinfo  {journal} {Angew. Chem. Int. Ed.}\ }\textbf
  {\bibinfo {volume} {55}},\ \bibinfo {pages} {6826} (\bibinfo {year}
  {2016})}\BibitemShut {NoStop}%
\bibitem [{\citenamefont {Simoncelli}\ \emph {et~al.}(2019)\citenamefont
  {Simoncelli}, \citenamefont {Marzari},\ and\ \citenamefont
  {Mauri}}]{simoncelli2019unified}%
  \BibitemOpen
  \bibfield  {author} {\bibinfo {author} {\bibfnamefont {M.}~\bibnamefont
  {Simoncelli}}, \bibinfo {author} {\bibfnamefont {N.}~\bibnamefont
  {Marzari}},\ and\ \bibinfo {author} {\bibfnamefont {F.}~\bibnamefont
  {Mauri}},\ }\bibfield  {title} {\bibinfo {title} {Unified theory of thermal
  transport in crystals and glasses},\ }\href
  {https://api.semanticscholar.org/CorpusID:256713150} {\bibfield  {journal}
  {\bibinfo  {journal} {Nat. Phys.}\ }\textbf {\bibinfo {volume} {15}},\
  \bibinfo {pages} {809} (\bibinfo {year} {2019})}\BibitemShut {NoStop}%
\bibitem [{\citenamefont {El-Sharkawy}\ \emph {et~al.}(1983)\citenamefont
  {El-Sharkawy}, \citenamefont {Abou El-Azm}, \citenamefont {Kenawy},
  \citenamefont {Hillal},\ and\ \citenamefont
  {Abu-Basha}}]{el1983thermophysical}%
  \BibitemOpen
  \bibfield  {author} {\bibinfo {author} {\bibfnamefont {A.}~\bibnamefont
  {El-Sharkawy}}, \bibinfo {author} {\bibfnamefont {A.}~\bibnamefont {Abou
  El-Azm}}, \bibinfo {author} {\bibfnamefont {M.}~\bibnamefont {Kenawy}},
  \bibinfo {author} {\bibfnamefont {A.}~\bibnamefont {Hillal}},\ and\ \bibinfo
  {author} {\bibfnamefont {H.}~\bibnamefont {Abu-Basha}},\ }\bibfield  {title}
  {\bibinfo {title} {Thermophysical properties of polycrystalline pbs, pbse,
  and pbte in the temperature range 300--700 k},\ }\href
  {https://api.semanticscholar.org/CorpusID:121079451} {\bibfield  {journal}
  {\bibinfo  {journal} {Int. J. Thermophys.}\ }\textbf {\bibinfo {volume}
  {4}},\ \bibinfo {pages} {261} (\bibinfo {year} {1983})}\BibitemShut {NoStop}%
\bibitem [{\citenamefont {Akhmedova}\ and\ \citenamefont
  {Abdinov}(2009)}]{akhmedova2009effect}%
  \BibitemOpen
  \bibfield  {author} {\bibinfo {author} {\bibfnamefont {G.}~\bibnamefont
  {Akhmedova}}\ and\ \bibinfo {author} {\bibfnamefont {D.~S.}\ \bibnamefont
  {Abdinov}},\ }\bibfield  {title} {\bibinfo {title} {Effect of thallium doping
  on the thermal conductivity of pbte single crystals},\ }\href
  {https://doi.org/10.1134/S0020168509080056} {\bibfield  {journal} {\bibinfo
  {journal} {Inorg. Mater.}\ }\textbf {\bibinfo {volume} {45}},\ \bibinfo
  {pages} {854} (\bibinfo {year} {2009})}\BibitemShut {NoStop}%
\bibitem [{\citenamefont {H{\aa}kansson}\ and\ \citenamefont
  {Ross}(1985)}]{haakansson1985thermal}%
  \BibitemOpen
  \bibfield  {author} {\bibinfo {author} {\bibfnamefont {B.}~\bibnamefont
  {H{\aa}kansson}}\ and\ \bibinfo {author} {\bibfnamefont {R.}~\bibnamefont
  {Ross}},\ }\bibfield  {title} {\bibinfo {title} {Thermal conductivity of
  solid naf under high pressure},\ }\href
  {https://doi.org/https://doi.org/10.1007/BF00500268} {\bibfield  {journal}
  {\bibinfo  {journal} {Int. J. Thermophys.}\ }\textbf {\bibinfo {volume}
  {6}},\ \bibinfo {pages} {353} (\bibinfo {year} {1985})}\BibitemShut {NoStop}%
\bibitem [{\citenamefont {H{\aa}kansson}\ and\ \citenamefont
  {Andersson}(1986)}]{haakansson1986thermal}%
  \BibitemOpen
  \bibfield  {author} {\bibinfo {author} {\bibfnamefont {B.}~\bibnamefont
  {H{\aa}kansson}}\ and\ \bibinfo {author} {\bibfnamefont {P.}~\bibnamefont
  {Andersson}},\ }\bibfield  {title} {\bibinfo {title} {Thermal conductivity
  and heat capacity of solid nacl and nai under pressure},\ }\href
  {https://api.semanticscholar.org/CorpusID:98218713} {\bibfield  {journal}
  {\bibinfo  {journal} {J Phys Chem Solids}\ }\textbf {\bibinfo {volume}
  {47}},\ \bibinfo {pages} {355} (\bibinfo {year} {1986})}\BibitemShut
  {NoStop}%
\bibitem [{\citenamefont {Steigmeier}\ and\ \citenamefont
  {Kudman}(1963)}]{steigmeier1963thermal}%
  \BibitemOpen
  \bibfield  {author} {\bibinfo {author} {\bibfnamefont {E.}~\bibnamefont
  {Steigmeier}}\ and\ \bibinfo {author} {\bibfnamefont {I.}~\bibnamefont
  {Kudman}},\ }\bibfield  {title} {\bibinfo {title} {Thermal conductivity of
  iii-v compounds at high temperatures},\ }\href
  {https://api.semanticscholar.org/CorpusID:121279431} {\bibfield  {journal}
  {\bibinfo  {journal} {Phys. Rev.}\ }\textbf {\bibinfo {volume} {132}},\
  \bibinfo {pages} {508} (\bibinfo {year} {1963})}\BibitemShut {NoStop}%
\bibitem [{\citenamefont {Madelung}(2004)}]{madelung2004semiconductors}%
  \BibitemOpen
  \bibfield  {author} {\bibinfo {author} {\bibfnamefont {O.}~\bibnamefont
  {Madelung}},\ }\href
  {https://link.springer.com/book/10.1007/978-3-642-18865-7} {\emph {\bibinfo
  {title} {Semiconductors: data handbook}}}\ (\bibinfo  {publisher} {Springer
  Science \& Business Media},\ \bibinfo {year} {2004})\BibitemShut {NoStop}%
\bibitem [{\citenamefont {Kang}\ \emph {et~al.}(2018)\citenamefont {Kang},
  \citenamefont {Li}, \citenamefont {Wu}, \citenamefont {Nguyen},\ and\
  \citenamefont {Hu}}]{kang2018experimental}%
  \BibitemOpen
  \bibfield  {author} {\bibinfo {author} {\bibfnamefont {J.~S.}\ \bibnamefont
  {Kang}}, \bibinfo {author} {\bibfnamefont {M.}~\bibnamefont {Li}}, \bibinfo
  {author} {\bibfnamefont {H.}~\bibnamefont {Wu}}, \bibinfo {author}
  {\bibfnamefont {H.}~\bibnamefont {Nguyen}},\ and\ \bibinfo {author}
  {\bibfnamefont {Y.}~\bibnamefont {Hu}},\ }\bibfield  {title} {\bibinfo
  {title} {Experimental observation of high thermal conductivity in boron
  arsenide},\ }\href {https://www.science.org/doi/pdf/10.1126/science.aat5522}
  {\bibfield  {journal} {\bibinfo  {journal} {Science}\ }\textbf {\bibinfo
  {volume} {361}},\ \bibinfo {pages} {575} (\bibinfo {year}
  {2018})}\BibitemShut {NoStop}%
\bibitem [{\citenamefont {Kumashiro}\ \emph {et~al.}(1989)\citenamefont
  {Kumashiro}, \citenamefont {Mitsuhashi}, \citenamefont {Okaya}, \citenamefont
  {Muta}, \citenamefont {Koshiro}, \citenamefont {Takahashi},\ and\
  \citenamefont {Mirabayashi}}]{kumashiro1989thermal}%
  \BibitemOpen
  \bibfield  {author} {\bibinfo {author} {\bibfnamefont {Y.}~\bibnamefont
  {Kumashiro}}, \bibinfo {author} {\bibfnamefont {T.}~\bibnamefont
  {Mitsuhashi}}, \bibinfo {author} {\bibfnamefont {S.}~\bibnamefont {Okaya}},
  \bibinfo {author} {\bibfnamefont {F.}~\bibnamefont {Muta}}, \bibinfo {author}
  {\bibfnamefont {T.}~\bibnamefont {Koshiro}}, \bibinfo {author} {\bibfnamefont
  {Y.}~\bibnamefont {Takahashi}},\ and\ \bibinfo {author} {\bibfnamefont
  {M.}~\bibnamefont {Mirabayashi}},\ }\bibfield  {title} {\bibinfo {title}
  {Thermal conductivity of a boron phosphide single-crystal wafer up to high
  temperature},\ }\href {https://api.semanticscholar.org/CorpusID:97024813}
  {\bibfield  {journal} {\bibinfo  {journal} {J. Appl. Phys.}\ }\textbf
  {\bibinfo {volume} {65}},\ \bibinfo {pages} {2147} (\bibinfo {year}
  {1989})}\BibitemShut {NoStop}%
\bibitem [{\citenamefont {Slack}(1972)}]{slack1972thermal}%
  \BibitemOpen
  \bibfield  {author} {\bibinfo {author} {\bibfnamefont {G.~A.}\ \bibnamefont
  {Slack}},\ }\bibfield  {title} {\bibinfo {title} {Thermal conductivity of
  ii-vi compounds and phonon scattering by fe 2+ impurities},\ }\href
  {https://doi.org/10.1103/PhysRevB.6.3791} {\bibfield  {journal} {\bibinfo
  {journal} {Phys. Rev. B}\ }\textbf {\bibinfo {volume} {6}},\ \bibinfo {pages}
  {3791} (\bibinfo {year} {1972})}\BibitemShut {NoStop}%
\bibitem [{\citenamefont {Chen}\ \emph {et~al.}(2020)\citenamefont {Chen},
  \citenamefont {Song}, \citenamefont {Ravichandran}, \citenamefont {Zheng},
  \citenamefont {Chen}, \citenamefont {Lee}, \citenamefont {Sun}, \citenamefont
  {Li}, \citenamefont {Gamage}, \citenamefont {Tian}, \citenamefont {Ding},
  \citenamefont {Song}, \citenamefont {Rai}, \citenamefont {Wu}, \citenamefont
  {Koirala}, \citenamefont {Schmidt}, \citenamefont {Watanabe}, \citenamefont
  {Lv}, \citenamefont {Ren}, \citenamefont {Shi}, \citenamefont {Cahill},
  \citenamefont {Taniguchi}, \citenamefont {Broido},\ and\ \citenamefont
  {Chen}}]{doi:10.1126/science.aaz6149}%
  \BibitemOpen
  \bibfield  {author} {\bibinfo {author} {\bibfnamefont {K.}~\bibnamefont
  {Chen}}, \bibinfo {author} {\bibfnamefont {B.}~\bibnamefont {Song}}, \bibinfo
  {author} {\bibfnamefont {N.~K.}\ \bibnamefont {Ravichandran}}, \bibinfo
  {author} {\bibfnamefont {Q.}~\bibnamefont {Zheng}}, \bibinfo {author}
  {\bibfnamefont {X.}~\bibnamefont {Chen}}, \bibinfo {author} {\bibfnamefont
  {H.}~\bibnamefont {Lee}}, \bibinfo {author} {\bibfnamefont {H.}~\bibnamefont
  {Sun}}, \bibinfo {author} {\bibfnamefont {S.}~\bibnamefont {Li}}, \bibinfo
  {author} {\bibfnamefont {G.~A. G.~U.}\ \bibnamefont {Gamage}}, \bibinfo
  {author} {\bibfnamefont {F.}~\bibnamefont {Tian}}, \bibinfo {author}
  {\bibfnamefont {Z.}~\bibnamefont {Ding}}, \bibinfo {author} {\bibfnamefont
  {Q.}~\bibnamefont {Song}}, \bibinfo {author} {\bibfnamefont {A.}~\bibnamefont
  {Rai}}, \bibinfo {author} {\bibfnamefont {H.}~\bibnamefont {Wu}}, \bibinfo
  {author} {\bibfnamefont {P.}~\bibnamefont {Koirala}}, \bibinfo {author}
  {\bibfnamefont {A.~J.}\ \bibnamefont {Schmidt}}, \bibinfo {author}
  {\bibfnamefont {K.}~\bibnamefont {Watanabe}}, \bibinfo {author}
  {\bibfnamefont {B.}~\bibnamefont {Lv}}, \bibinfo {author} {\bibfnamefont
  {Z.}~\bibnamefont {Ren}}, \bibinfo {author} {\bibfnamefont {L.}~\bibnamefont
  {Shi}}, \bibinfo {author} {\bibfnamefont {D.~G.}\ \bibnamefont {Cahill}},
  \bibinfo {author} {\bibfnamefont {T.}~\bibnamefont {Taniguchi}}, \bibinfo
  {author} {\bibfnamefont {D.}~\bibnamefont {Broido}},\ and\ \bibinfo {author}
  {\bibfnamefont {G.}~\bibnamefont {Chen}},\ }\bibfield  {title} {\bibinfo
  {title} {Ultrahigh thermal conductivity in isotope-enriched cubic boron
  nitride},\ }\href {https://doi.org/10.1126/science.aaz6149} {\bibfield
  {journal} {\bibinfo  {journal} {Science}\ }\textbf {\bibinfo {volume}
  {367}},\ \bibinfo {pages} {555} (\bibinfo {year} {2020})}\BibitemShut
  {NoStop}%
\bibitem [{\citenamefont {Li}\ and\ \citenamefont
  {Mingo}(2014)}]{PhysRevB.89.184304}%
  \BibitemOpen
  \bibfield  {author} {\bibinfo {author} {\bibfnamefont {W.}~\bibnamefont
  {Li}}\ and\ \bibinfo {author} {\bibfnamefont {N.}~\bibnamefont {Mingo}},\
  }\bibfield  {title} {\bibinfo {title} {Thermal conductivity of fully filled
  skutterudites: Role of the filler},\ }\href
  {https://doi.org/10.1103/PhysRevB.89.184304} {\bibfield  {journal} {\bibinfo
  {journal} {Phys. Rev. B}\ }\textbf {\bibinfo {volume} {89}},\ \bibinfo
  {pages} {184304} (\bibinfo {year} {2014})}\BibitemShut {NoStop}%
\bibitem [{\citenamefont {He}\ \emph {et~al.}(2022)\citenamefont {He},
  \citenamefont {Xia}, \citenamefont {Lin}, \citenamefont {Pal}, \citenamefont
  {Zhu}, \citenamefont {Kanatzidis},\ and\ \citenamefont
  {Wolverton}}]{https://doi.org/10.1002/adfm.202108532}%
  \BibitemOpen
  \bibfield  {author} {\bibinfo {author} {\bibfnamefont {J.}~\bibnamefont
  {He}}, \bibinfo {author} {\bibfnamefont {Y.}~\bibnamefont {Xia}}, \bibinfo
  {author} {\bibfnamefont {W.}~\bibnamefont {Lin}}, \bibinfo {author}
  {\bibfnamefont {K.}~\bibnamefont {Pal}}, \bibinfo {author} {\bibfnamefont
  {Y.}~\bibnamefont {Zhu}}, \bibinfo {author} {\bibfnamefont {M.~G.}\
  \bibnamefont {Kanatzidis}},\ and\ \bibinfo {author} {\bibfnamefont
  {C.}~\bibnamefont {Wolverton}},\ }\bibfield  {title} {\bibinfo {title}
  {Accelerated discovery and design of ultralow lattice thermal conductivity
  materials using chemical bonding principles},\ }\href
  {https://doi.org/https://doi.org/10.1002/adfm.202108532} {\bibfield
  {journal} {\bibinfo  {journal} {Adv. Funct. Mater.}\ }\textbf {\bibinfo
  {volume} {32}},\ \bibinfo {pages} {2108532} (\bibinfo {year}
  {2022})}\BibitemShut {NoStop}%
\end{thebibliography}%

\end{document}